\documentclass[english,aps,prb,notitlepage,twocolumn,superscriptaddress,floatfix, 10pt]{revtex4-2}
\usepackage{CJK}
\usepackage{graphicx}
\usepackage{dcolumn}
\usepackage{bm}
\usepackage{mathtools}
\usepackage{amssymb,amsmath,amsfonts,dsfont}
\usepackage{hyperref}
\usepackage{color}

\newcommand{\tr}{\operatorname{Tr}}
\def\avg#1{\mathinner{\langle{#1}\rangle}}
\def\bra#1{\mathinner{\langle{#1}|}}
\def\ket#1{\mathinner{|{#1}\rangle}}

\newcommand{\proj}[1]{\ket{#1}\!\!\bra{#1}}

\newcommand{\norm}[1]{\left\lVert#1\right\rVert}
\newcommand{\su}{\text{SU}}

\begin{document}
\begin{CJK*}{UTF8}{mj}
    
\title{Simulating quantum dynamics in two-dimensional lattices with tensor network influence functional belief propagation}

\author{Gunhee Park (박건희)}
\email{gppark@caltech.edu}
\affiliation{Division of Engineering and Applied Science, California Institute of Technology, Pasadena, California 91125, USA}

\author{Johnnie Gray}
\affiliation{Division of Chemistry and Chemical Engineering, California Institute of Technology, Pasadena, California 91125, USA}

\author{Garnet Kin-Lic Chan}
\email{gkc1000@gmail.com}
\affiliation{Division of Chemistry and Chemical Engineering, California Institute of Technology, Pasadena, California 91125, USA}

\begin{abstract}
    Describing nonequilibrium quantum dynamics remains a significant computational challenge due to the growth of spatial entanglement. The tensor network influence functional (TN-IF) approach mitigates this problem for computing the time evolution of local observables
    by encoding the subsystem's influence functional path integral as a matrix product state (MPS), thereby shifting the resource governing computational cost from spatial entanglement to temporal entanglement.
    We extend the applicability of the TN-IF method to two-dimensional lattices by demonstrating its construction on tree lattices and proposing a belief propagation (BP) algorithm for the TN-IF, termed influence functional BP (IF-BP), to simulate local observable dynamics on arbitrary graphs. Even though the BP algorithm introduces uncontrolled approximation errors on arbitrary graphs, it provides an accurate description for locally tree-like lattices. Numerical simulations of the kicked Ising model on a heavy-hex lattice, motivated by a recent quantum experiment, highlight the effectiveness of the IF-BP method, which demonstrates superior performance in capturing long-time dynamics where traditional tensor network state-based methods struggle. Our results further reveal that the temporal entanglement entropy (TEE) only grows logarithmically with time for this model, resulting in a polynomial computational cost for the whole method.
    We further construct a cluster expansion of IF-BP to introduce loop correlations beyond the BP approximation, providing a systematic correction to the IF-BP estimate. We demonstrate the power of the cluster expansion of the IF-BP in simulating the quantum quench dynamics of the 2D transverse field Ising model, obtaining numerical results that improve on the state-of-the-art.
\end{abstract}

\maketitle
\end{CJK*}

\maketitle

\section{Introduction}

Describing nonequilibrium quantum dynamics is a formidable challenge, even in one-dimensional systems, due to the exponential growth of the Hilbert space with system size. 
For equilibrium properties, the area-law scaling of spatial entanglement allows for the efficient representation of ground states using tensor network (TN) states~\cite{CiracRMP2021}. In nonequilibrium settings, however, the situation is markedly different. In generic thermalizing systems, the entanglement entropy typically grows linearly in time, even when starting from an initially weakly entangled state~\cite{Calabrese_2005, KimHuse2013}. This growth results in an exponential increase in computational cost for simulating quantum dynamics with TN states.

To circumvent the exponential cost, numerous methods have been proposed to simulate the dynamics of local observables without requiring a full representation of the quantum state~\cite{White2018, Tamascelli2018, Surace2019, Perez2024convert, Rakovszky2022, Keyserlingk2022backflow, Kvorning2022, Artiaco2024, ParkHuang2024}. Among these, we focus on the tensor network influence functional (TN-IF) approach~\cite{Banuls2009, Hastings2015, Strathearn2018, Jorgensen2019, Ye2021, Lerose2021, SONNER2021168677, Friasperez2022lightcone, Foligno2023, Cygorek2022, NathanNg2023, park2024tensor, Thoenniss2023, ChenGuo2024, Nguyen2024, Fux2024oqupy}. The influence functional (IF) concept originates from the open quantum system description of quantum dynamics~\cite{feynman1963}, where the system is divided into a subsystem and a bath. The subsystem contains the degrees of freedom associated with the observable of interest, while the IF captures the time-nonlocal influence of the bath through a path integral over the subsystem's trajectories after tracing out the bath. Recent studies have demonstrated that the IF can often be efficiently represented as a compact, low-rank matrix product state (MPS) along the time direction. Most prior work has focused on IFs derived from Gaussian baths~\cite{Strathearn2018, Jorgensen2019, Ye2021, NathanNg2023, park2024tensor, Thoenniss2023, ChenGuo2024, Nguyen2024, Fux2024oqupy}, one-dimensional (1D) interacting baths~\cite{Banuls2009, Hastings2015, Ye2021, Lerose2021, SONNER2021168677, Friasperez2022lightcone, Foligno2023}, or separable interacting baths~\cite{Cygorek2022}.

This paper aims to extend the TN-IF method's applicability to two-dimensional (2D) lattice models. To address the complexity inherent to 2D systems, we begin by constructing the TN-IF for dynamics on tree lattices. The absence of loops in tree lattices ensures that the structure of the IF closely resembles that found in 1D systems. Building on this tree construction, we introduce a belief propagation (BP) algorithm adapted to the IF, which we refer to as influence functional BP (IF-BP), for simulating dynamics on arbitrary graphs. The BP algorithm, originally developed for probabilistic graphical models~\cite{Yedidia2000, yedidia2000belief, Mezard_Montanari_book_2009}, has recently been explored in the context of general tensor network contractions~\cite{Alkabetz2021, sahu2022efficient, pancotti2023onestep, Tindall2023gauging, Guo2023blockBP, Begusic2024, tindall2025dynamicsdisordered}. While the BP algorithm introduces an uncontrolled approximation, it often achieves reasonable accuracy on locally tree-like lattices with large loops.

The efficacy of the proposed approach is illustrated through numerical simulations of the kicked Ising model on a heavy-hex lattice, inspired by recent quantum experiments with IBM's superconducting qubit processor~\cite{Kim2023evidence}. The large-loop structure of the heavy-hex lattice facilitates the application of TN states and TN operators with BP approximations. These methods have shown high accuracy in classical simulations, successfully capturing the dynamics up to the experimental timescale of 20 Trotter steps~\cite{Begusic2024, Tindall2024ibm, liao2023simulation, Orus2024ibm}. Nonetheless, while BP-approximated TN states perform well at short times, their applicability is ultimately constrained by the growth of spatial entanglement, which can result in exponential computational costs for longer-time simulations.

\begin{figure*}[t]
    \centering
    \includegraphics[width=0.8\textwidth]{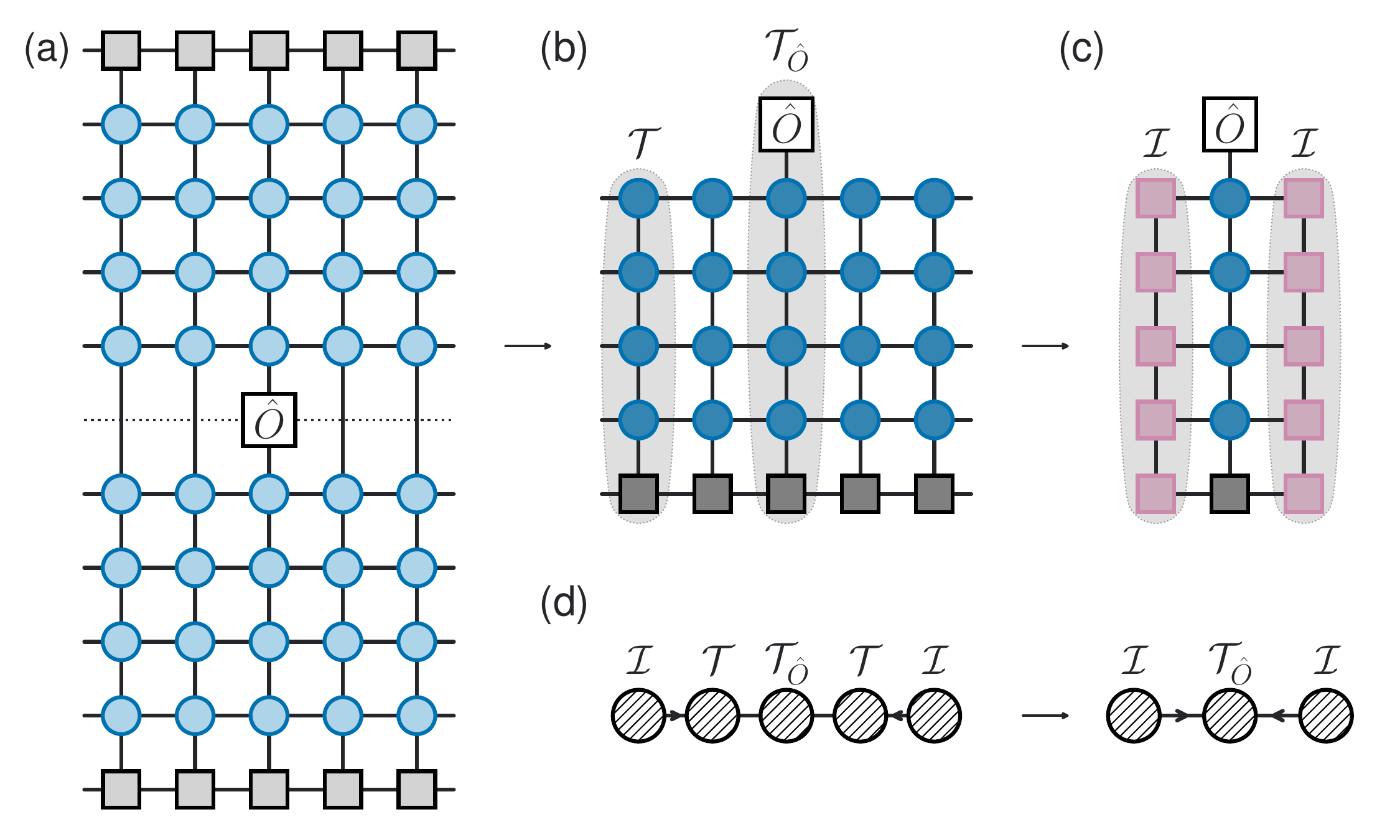}
    \caption{Tensor network diagram illustrating dynamics on a one-dimensional lattice. (a) Tensor network diagram for a local observable expectation value $\langle \hat{O}(m) \rangle = \bra{\psi(m)} \hat{O} \ket{\psi(m)}$ where $m=4$ in this figure. The initial state $\ket{\psi_0}$ is given by a matrix product state (MPS) (light gray squares), and time evolution is carried out by a matrix product operator (MPO) (light blue circles). (b) A folded tensor network (folded TN), where the folding occurs along the dashed line in (a). The darker shaded tensors denote doubly grouped tensors from (a). Tensors from the same spatial site are grouped together as $\mathcal{T}$ or $\mathcal{T}_{\hat{O}}$, shown with the light gray shading. (c) We contract the folded TN in (b) along the transverse direction around the site with $\hat{O}$. The boundary tensors are approximated by an MPS with a fixed bond dimension during the transverse contraction. The final MPS $\mathcal{I}$ (pink squares) is called an influence functional MPS (IF-MPS). (d) One-dimensional structure of the IF-MPS propagation after grouping tensors by spatial sites. }
    \label{fig:IFMPS_1d}
\end{figure*}

Numerical results using the IF-BP show enhanced classical simulation performance already at 20 Trotter steps for specific parameter regimes. For longer-time dynamics, the IF-BP successfully captures the magnetization of an infinite-temperature state in agreement with predictions from Floquet thermalization~\cite{Lazarides2014, DAlessio2014, PONTE2015196}. In contrast, the time-propagated TN states with finite bond dimension exhibit an unphysical increase in magnetization. The computational advantage of the IF-BP is attributed to the slow growth of the temporal entanglement entropy~\cite{Lerose2021, SONNER2021168677}, which here increases \emph{logarithmically} with time. The resulting logarithmic scaling suggests that the MPS representation of the IF requires only a polynomial computational cost.

We further extend the TN-IF framework beyond the BP approximation to account for loop effects in general loopy graphs, such as square lattices, by employing a cluster expansion. This approach allows us to systematically correct the uncontrolled approximations inherent in the BP limit, improving accuracy for graphs with significant loop contributions. To demonstrate the effectiveness of this extension, we analyze the quench dynamics of the transverse field Ising model on a square lattice - a case where the BP solution alone fails to provide accurate results due to the pronounced influence of loops. Comparisons with several methods, including TN state propagation, sparse Pauli dynamics~\cite{Begusic2024realtime}, and neural quantum Galerkin methods~\cite{Sinibaldi2024}, demonstrate the state-of-the-art accuracy of our approach.

\section{Tensor Network Influence functionals on tree lattices}\label{sec:sec2}

\begin{figure*}[t]
    \centering
    \includegraphics[width=0.9\textwidth]{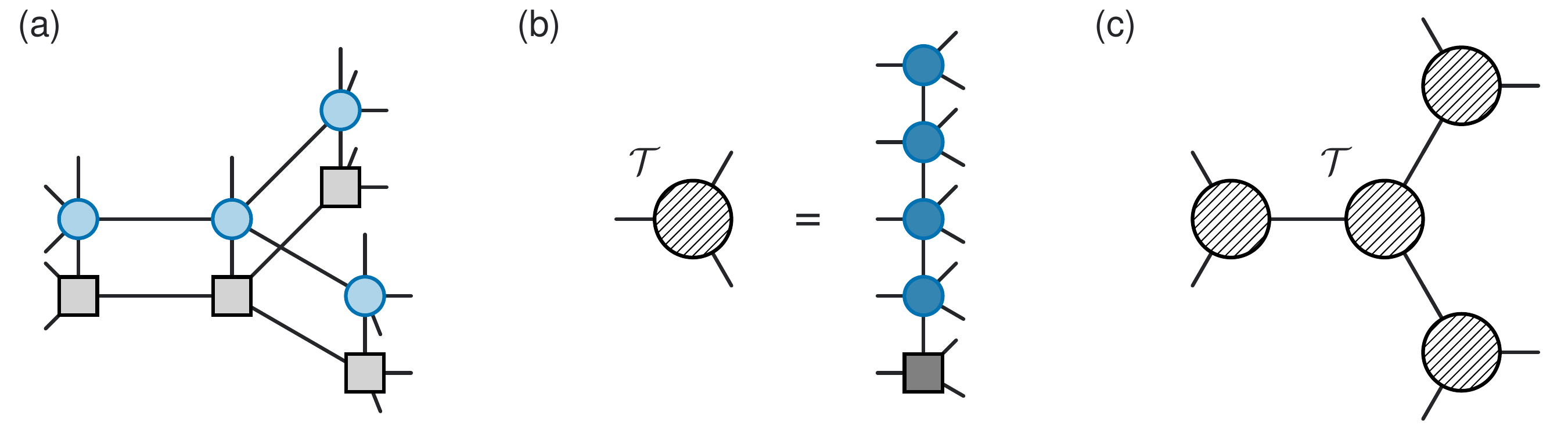}
    \caption{Tensor network diagram illustrating dynamics on a tree lattice. (a) The initial state is a tree tensor network state, and its time evolution is carried out by a tree tensor network operator. (b) After folding the TN $\langle \hat{O}(m) \rangle$, we group tensors in the same spatial site as $\mathcal{T}$. (c) A grouped tensor $\mathcal{T}$ within
     a tree tensor network. }
    \label{fig:tree1}
\end{figure*}

\begin{figure*}[t]
    \centering
    \includegraphics[width=1.0\textwidth]{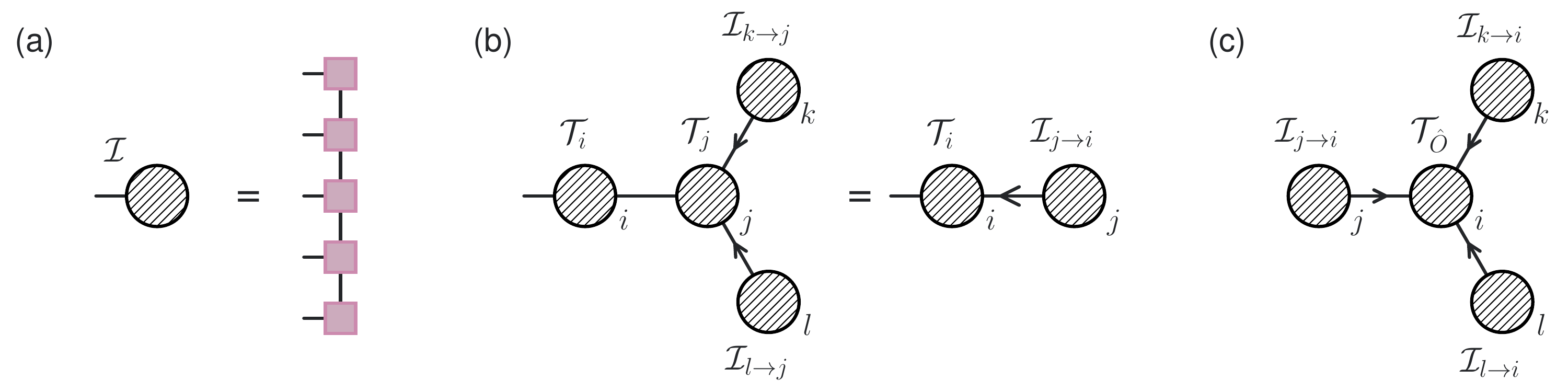}
    \caption{(a) IF-MPS $\mathcal{I}$ for the tree subgraph at each bond. (b) TN contraction of two IF-MPSs $\mathcal{I}_{k \rightarrow j}$ and $\mathcal{I}_{l \rightarrow j}$ from sites $k$ and $l$ with the tensor $\mathcal{T}_j$ at site $j$. This creates a new IF-MPS $\mathcal{I}_{j \rightarrow i}$ directed from site $j$ to site $i$. (c) Final TN diagram after the IF-MPSs reach the site of interest with the operator $\hat{O}$. The desired expectation value is computed from the contraction of this TN. }
    \label{fig:tree2}
\end{figure*}

\subsection{Dynamics on one-dimensional lattices}

In this section, we formulate tensor network influence functionals on tree lattices. Before considering the dynamics on tree lattices, we first review tensor network influence functionals on 1D lattices~\cite{Banuls2009, Lerose2021, SONNER2021168677, Ye2021}, which are the simplest type of tree lattice. We consider a time evolution of a 1D system starting from a matrix product state (MPS) $\ket{\psi_0}$. We assume time evolution is described by a unitary matrix product operator (MPO)~\cite{Pirvu_2010}, denoted as $\hat{U}$. A time-evolved wavefunction after $m$ applications of the MPO can be written as $\ket{\psi(m)} = \hat{U}_{m} \cdots \hat{U}_2 \hat{U}_1 \ket{\psi_0} $. Our goal is to compute an expectation value of a local single-site observable $\hat{O}$,  $\avg{\hat{O}(m)} = \bra{\psi(m)} \hat{O} \ket{\psi(m)}$. A tensor network (TN) diagram representing this expectation value is illustrated in Fig.~\ref{fig:IFMPS_1d}a, and has a two-dimensional (2D) TN structure.

Computing the expectation value from the TN diagram requires contracting the 2D space-time TN. The accuracy and computational cost of 2D TN contraction depend on the TN contraction path and approximation scheme. One standard way of contracting a 2D TN is to approximate the boundary of the 2D TN with an MPS~\cite{verstraete2004renormalization, VerstraeteMurgCirac2008, Lubasch2014prb, Lubasch2014}.
The standard contraction path 
represents $\ket{\psi(m)}$ as a fixed bond-dimension MPS and contracts the time evolution along the time direction. 

Ref.~\cite{Banuls2009} proposed a new strategy for contracting the space-time TN along the spatial direction, referring to this as a transverse contraction. One significant observation is that the accuracy of the transverse contraction improves after folding the TN in half and grouping the ket and bra parts ($\ket{\psi(m)}$ and $\bra{\psi(m)}$, respectively) as in Fig.~\ref{fig:IFMPS_1d}b. The folded TN is contracted transversely towards the column with $\hat{O}$ ($\mathcal{T}_{\hat{O}}$ in Fig.~\ref{fig:IFMPS_1d}b), and the intermediate boundary tensors are approximated with MPS with a fixed bond dimension. After contracting all the tensors on the left and right sides of $\mathcal{T}_{\hat{O}}$, we obtain two MPSs, denoted by $\mathcal{I}$ and pink tensors in Fig.~\ref{fig:IFMPS_1d}c. The expectation value of the local observable $\hat{O}$ is computed by an overlap of the two $\mathcal{I}$ with the MPO $\mathcal{T}_{\hat{O}}$ in the middle.

In recent years, the above folding strategy has been re-interpreted in the language of tensor network influence functionals (TN-IF)~\cite{Lerose2021, SONNER2021168677, Ye2021}. The folded TN in Fig.~\ref{fig:IFMPS_1d}b can be expressed as $\tr[\hat{O} \mathcal{U}_m \cdots \mathcal{U}_2 \mathcal{U}_1 \hat{\rho}_0]$ where $\hat{\rho}_0 = \proj{\psi_0}$ and the superoperator $\mathcal{U}$ is defined by $\mathcal{U} \bullet = \hat{U} \bullet \hat{U}^\dag$. The sites outside of $\hat{O}$ are viewed as a bath, and the transverse contraction effectively traces out the bath, only leaving its influence on the dynamics of the subsystem. This effect of a bath on a system's dynamics was originally formulated by Feynman and Vernon~\cite{feynman1963} in terms of an influence functional (IF) path integral over the system density operator degrees of freedom for baths consisting of harmonic oscillators linearly coupled to the system. The contracted MPS $\mathcal{I}$ defines an influence functional~\footnote{More precisely, $\mathcal{I}$ reweights the path integral over bond degrees of freedom in Fig.~\ref{fig:IFMPS_1d}b, not over the system degrees of freedom. This object is defined as a boundary influence functional in \cite{park2024tensor}. In this work, we do not distinguish between the two. } for arbitrary many-body quantum baths and couplings.
We describe this object as an influence functional MPS (IF-MPS).

The IF-MPS approach provides a controlled way to approximate the local expectation value by increasing the bond dimension of the IF-MPS. The IF-MPS $\mathcal{I}$ can be obtained in a finite system by propagating (contracting) the tensor network in the spatial direction starting from a spatial boundary. In a translationally invariant infinite system, we can further formulate an equation to find the leading eigenvector of the transfer matrix $\mathcal{T}$ in Fig.~\ref{fig:IFMPS_1d}b and \ref{fig:IFMPS_1d}d, which then serves as an effective spatial boundary.

The contraction of the IF-MPS along the spatial direction reduces to a 1D TN contraction of grouped tensors, where each grouped tensor corresponds to a column of tensors associated with a single spatial site (Fig.~\ref{fig:IFMPS_1d}d). Since arbitrary 1D TN can be contracted exactly, and the resulting effective tensor contraction can be efficiently carried out using the boundary MPS procedure, this provides a practical computational algorithm. Building on this approach, we next extend the formulation of TN-IF to tree lattices.

\subsection{Dynamics on tree lattices}

Now we consider dynamics on tree lattices. We assume that the initial state is a tree tensor network state, and a time evolution step is carried out by a tensor network operator with the same connectivity as the tree lattice, i.e., a tree tensor network operator, $\hat{U}$ in Fig.~\ref{fig:tree1}a. The folded TN for $\avg{\hat{O}(m)} = \bra{\psi(m)} \hat{O} \ket{\psi(m)}$ is constructed analogously to in 1D lattices. By grouping the column of tensors on the same site as $\mathcal{T}$ (Fig.~\ref{fig:tree1}b), we have a tree tensor network composed of the grouped tensors $\mathcal{T}$ (Fig.~\ref{fig:tree1}c).

\begin{figure}[t]
    \centering
    \includegraphics[width=0.75\columnwidth]{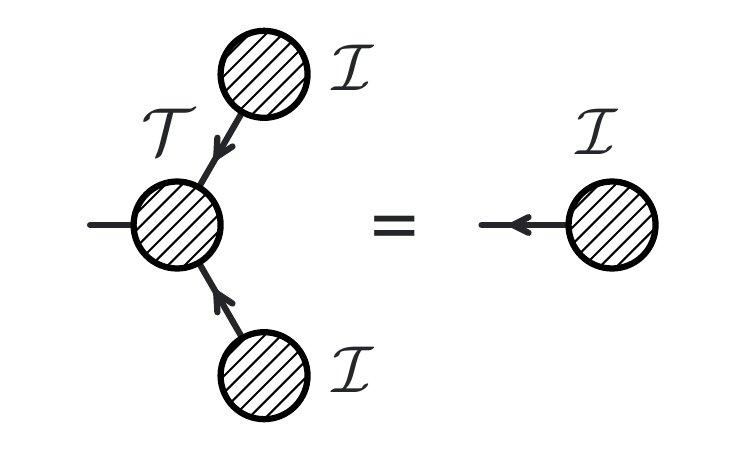}
    \caption{Self-consistent equation for IF-MPS in the $z=3$ Bethe lattice assuming rotational invariance.}
    \label{fig:bethe}
\end{figure}

Like in the 1D case, this tree tensor network has an efficient contraction path. Therefore, we can use the boundary contraction algorithm in conjunction with this path to contract the network, 
similar to the strategy used in the 1D lattice dynamics discussed above. In the previous section, we also related the boundary MPS contraction of a folded TN to the IF-MPS. Here, we extend the concept of IF-MPS to tree lattices.

At each bond extending from a site in the tree lattice, the subgraph from the bond also forms a tree disconnected from the tree subgraphs formed by the other bonds. For finite tree lattices, each subgraph is terminated by
end-sites that only have a single bond; for such an end-site, the corresponding tensor $\mathcal{I}$ has an MPS structure along the time axis (Fig.~\ref{fig:tree2}a), which defines an initial IF-MPS for the end-site. The main difference between 1D and tree lattice dynamics is that as we contract inwards from the end-sites, multiple IF-MPSs are combined at the same site. For example, in Fig.~\ref{fig:tree2}b, two sites $k$ and $l$ are connected around site $j$ with tensor $\mathcal{T}_j$, with IF-MPS $\mathcal{I}_{k \rightarrow j}$ and $\mathcal{I}_{l \rightarrow j}$. Here, we draw arrows on the bonds and subscripts to indicate the propagation direction more clearly. After contracting $\mathcal{T}_j$, $\mathcal{I}_{k \rightarrow j}$, and $\mathcal{I}_{l \rightarrow j}$, the new IF-MPS $\mathcal{I}_{j \rightarrow i}$ is created directed towards the site $i$. The bond dimension of $\mathcal{I}_{j \rightarrow i}$ can be truncated to a smaller finite bond dimension using MPS compression. In a  later section, we will discuss a more efficient numerical scheme to directly construct $\mathcal{I}_{j \rightarrow i}$ with a given bond dimension. We iterate this propagation until we reach the site of interest with $\hat{O}$, and the desired expectation value is computed from the contraction of the final TN as in Fig.~\ref{fig:tree2}c.

It is straightforward to extend the above procedure for finite tree lattices to infinite tree lattices with the same coordination number, known as Bethe lattices~\cite{Bethe1935}. Given the IF-MPS $\mathcal{I}$ for one tree subgraph, the subgraph itself is composed of the tensor $\mathcal{T}$ and $z-1$ IF-MPS $\mathcal{I}$ where $z$ is the coordination number of the Bethe lattice. Fig.~\ref{fig:bethe} illustrates the case when $z=3$. From the above condition, we see that the IF-MPS satisfies a non-linear self-consistent equation.

\section{Belief propagation with influence functionals}\label{sec:sec3}

\begin{figure}[t]
    \centering
    \includegraphics[width=\columnwidth]{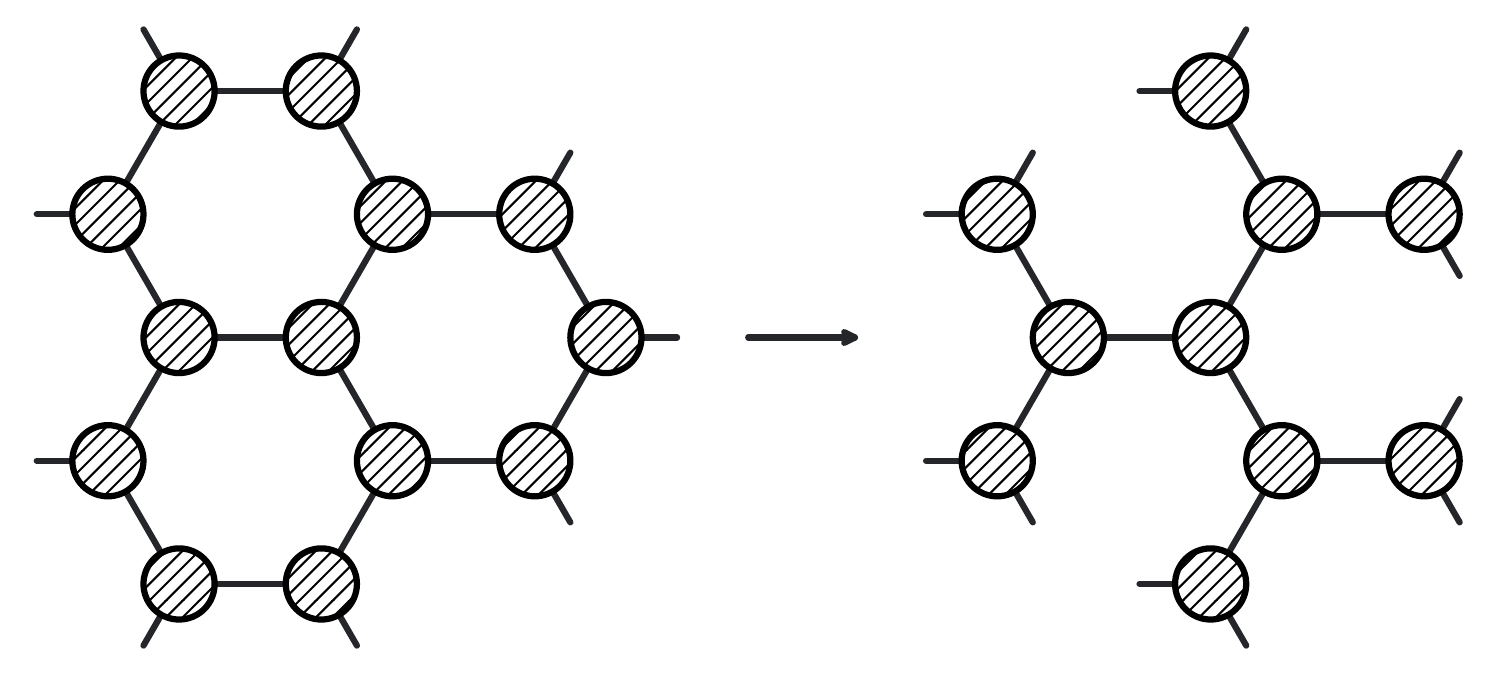}
    \caption{(Left) Tensor network diagram on an infinite hexagonal lattice. (Right) Within the belief propagation (BP) approximation, the fixed point BP equation is equivalent to that on the $z=3$ Bethe lattice.}
    \label{fig:bphex}
\end{figure}

This section introduces a belief propagation (BP) algorithm for IF-MPS, which we call influence functional belief propagation (IF-BP). IF-BP describes a numerical scheme to find the (self-consistent) IF-MPS $\mathcal{I}$ and to evaluate local expectation values using the IF-MPS. This algorithm applies to both tree and locally tree-like lattices. When the underlying graph is a tree, the accuracy of expectation values computed using IF-BP is controlled only by the bond dimension of the IF-MPS. When the graph is not exactly a tree, the IF-BP expectation values are not exact anymore, even in the infinite bond dimension limit. Nonetheless, the BP algorithm often provides good heuristic estimates when the graph is locally tree-like~\cite{Mezard_Montanari_book_2009}.

We start with the folded TN, where we group tensors by their sites as in Fig.~\ref{fig:tree1}c, but within a general lattice with loops.
IF-MPSs are initialized at each bond for both directions. For example, two IF-MPSs $\mathcal{I}_{i \rightarrow j}^{(0)}$ and $\mathcal{I}_{j \rightarrow i}^{(0)}$ are initialized at the bond between sites $i$ and $j$. The IF-BP involves iterative local propagation of the IF-MPS, similar to the propagation in Fig.~\ref{fig:tree2}b,
\begin{equation}
    \mathcal{I}_{j \rightarrow i}^{(t+1)} = \mathcal{T}_j \cdot \bigotimes_{k \in \partial j \backslash i} \mathcal{I}_{k \rightarrow j}^{(t)},
    \label{eq:ifbp}
\end{equation}
where $\partial j \backslash i$ denotes a set of neighbor sites of $j$ excluding site $i$ and $\cdot$ denotes a tensor contraction. We repeat the propagation until  Eq.~\ref{eq:ifbp} reaches a fixed point; it forms a local self-consistent BP equation~\cite{yedidia2000belief, Mezard_Montanari_book_2009},
\begin{equation}
    \mathcal{I}_{j \rightarrow i} = \mathcal{T}_j \cdot \bigotimes_{k \in \partial j \backslash i} \mathcal{I}_{k \rightarrow j}.
    \label{eq:ifbp2}
\end{equation}
The BP equation in the tree graph reduces to the exact propagation described in the previous section.

We remark that the BP equation introduced in this section resembles the BP equation for probabilistic graphical models~\cite{Mezard_Montanari_book_2009, yedidia2000belief} or tensor networks~\cite{Alkabetz2021, pancotti2023onestep, Tindall2023gauging, sahu2022efficient}. The BP equation in the above references is based on a `message-passing' algorithm where the message plays the role of the IF-MPS in this work. Unlike the standard message-passing algorithm, the main difference here is that the IF-MPS is approximated as a low-rank MPS with a fixed bond dimension during propagation~\footnote{There exists an extension of BP to approximate messages using MPS~\cite{Guo2023blockBP, tindall2025dynamicsdisordered}. However, there is a difference in the way the tensors are grouped. For example, in \cite{Guo2023blockBP}, the main goal is to compute the contraction of a 2D tensor network. Spatial sites are grouped into blocks, and the messages around the blocks are approximated as MPS.
In contrast, the main goal of our work is to contract $n+1$-dimensional tensor networks (where $n$ is the spatial dimension) in the context of quantum dynamics. A group of tensors is formed around a single site (or cluster of sites), but the message along the time direction is approximated as the MPS.}.

\begin{figure}[t]
    \centering
    \includegraphics[width=\columnwidth]{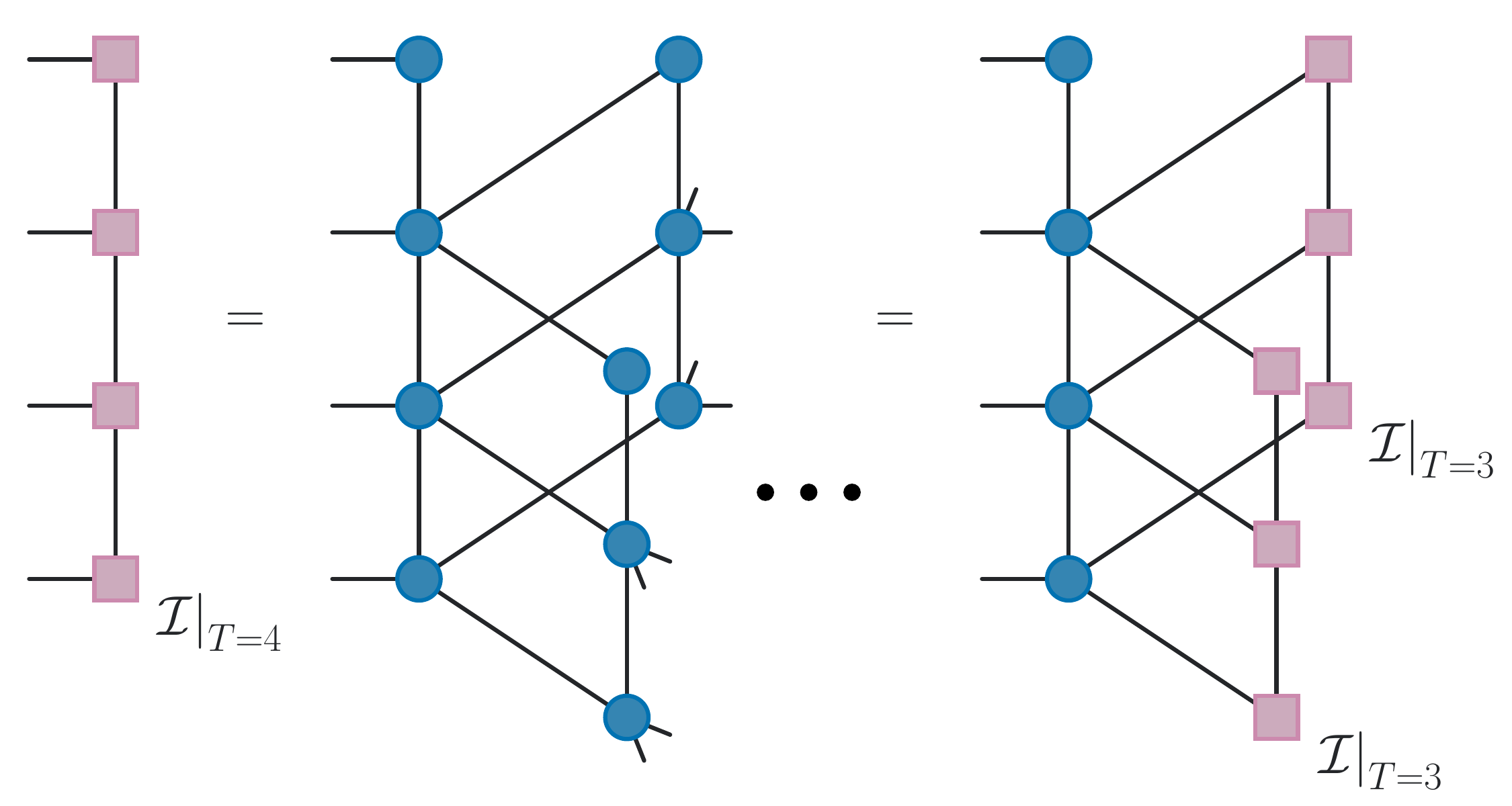}
    \caption{The IF-MPS from the Bethe lattice has a light cone structure. The light cone IF-MPS follows an iterative structure, where the IF-MPS at $T=4$, $\left.\mathcal{I}\right\vert_{T=4}$, can be constructed from two IF-MPSs at $T=3$, $\left.\mathcal{I}\right\vert_{T=3}$.}
    \label{fig:lightcone}
\end{figure}

The BP approximation becomes clearer when considering the dynamics of an infinite translation-invariant lattice, such as a hexagonal lattice in Fig.~\ref{fig:bphex}. In this limit, if the initialized IF-MPSs in each bond are invariant with respect to translations and rotations, the fixed point BP equation for the hexagonal lattice becomes identical to the self-consistent equation for the $z=3$ Bethe lattice in Fig.~\ref{fig:bethe}. Therefore, a BP approximated dynamics on the infinite hexagonal lattice yields a $z=3$ Bethe lattice dynamics (Fig.~\ref{fig:bphex}).

Using the Bethe lattice reduction, the fixed point IF-MPS on the infinite lattice can be directly obtained from a light cone on the Bethe lattice for the unitary dynamics. As illustrated in Fig.~\ref{fig:lightcone}, the light cone IF-MPS follows an iterative structure, where the IF-MPS at $T=4$, $\left.\mathcal{I}\right\vert_{T=4}$, can be constructed from two IF-MPSs at $T=3$, $\left.\mathcal{I}\right\vert_{T=3}$,
\begin{equation}
    \left.\mathcal{I}_{j \rightarrow i}\right\vert_{T=4} = \left.\mathcal{T}_j\right\vert_{T=4} \cdot \bigotimes_{k \in \partial j \backslash i}  \left.\mathcal{I}_{k \rightarrow j}\right\vert_{T=3}. \label{eq:lightcone}
\end{equation}
This light cone construction for the IF-MPS offers additional numerical advantages, even in 1D systems~\cite{Lerose2023overcome, Friasperez2022lightcone}, as the intermediate IF-MPS exhibits lower entanglement entropy.

Either with the IF-BP equation in Eq.~\ref{eq:ifbp} or with the light cone propagation in Eq.~\ref{eq:lightcone}, the bond dimension of the IF-MPS increases during the propagation, so bond dimension truncation during the propagation is necessary. An iterative singular value decomposition (SVD) is a standard way to truncate MPS, but it becomes expensive at large $z$. If the bond dimension of $\mathcal{I}^{(t)}_{k \rightarrow j}$ and $\mathcal{T}_j$ is $D$ and $d$, respectively, the bond dimension of $\mathcal{I}^{(t+1)}_{j \rightarrow i}$ without truncation is $dD^{z-1}$ for coordination number $z$. The computational cost of the SVD becomes $\mathcal{O}(D^{3(z-1)})$. Instead, we construct $\mathcal{I}^{(t+1)}_{j \rightarrow i}$ with a lower bond dimension directly from a variational optimization by minimizing the distance,
\begin{equation}
    d\left(\mathcal{I}^{(t+1)}_{j \rightarrow i}\right) = \norm{\mathcal{I}^{(t+1)}_{j \rightarrow i} - \mathcal{T}_j \cdot \bigotimes_k  \mathcal{I}^{(t)}_{k \rightarrow j} }^2 .
\end{equation}
The optimal solution can be found by an alternating least squares (ALS) optimization. This performs the least squares minimization for one tensor in $\mathcal{I}^{(t+1)}_{j \rightarrow i}$ at a time with the other tensors fixed and sweeps across the tensors in $\mathcal{I}^{(t+1)}_{j \rightarrow i}$ iteratively. This method has previously been utilized for MPO-MPS multiplication~\cite{verstraete2004renormalization, Stoudenmire_2010, Lubasch2014prb, Lubasch2014, PAECKEL2019167998}.

The leading computational cost of the ALS method depends on the computation of an overlap tensor network contraction, 
\begin{equation}
    \mathcal{I}^{(t+1)*}_{j \rightarrow i} \cdot \mathcal{T}_j \cdot \bigotimes_k  \mathcal{I}^{(t)}_{k \rightarrow j}.
\end{equation}
The computational complexity of this tensor network contraction is $\mathcal{O}(D^{z+1})$, lower than that of the SVD truncation for $z>2$.

\section{Perfect dephaser point}\label{sec:sec4}

\begin{figure}[t]
    \centering
    \includegraphics[width=\columnwidth]{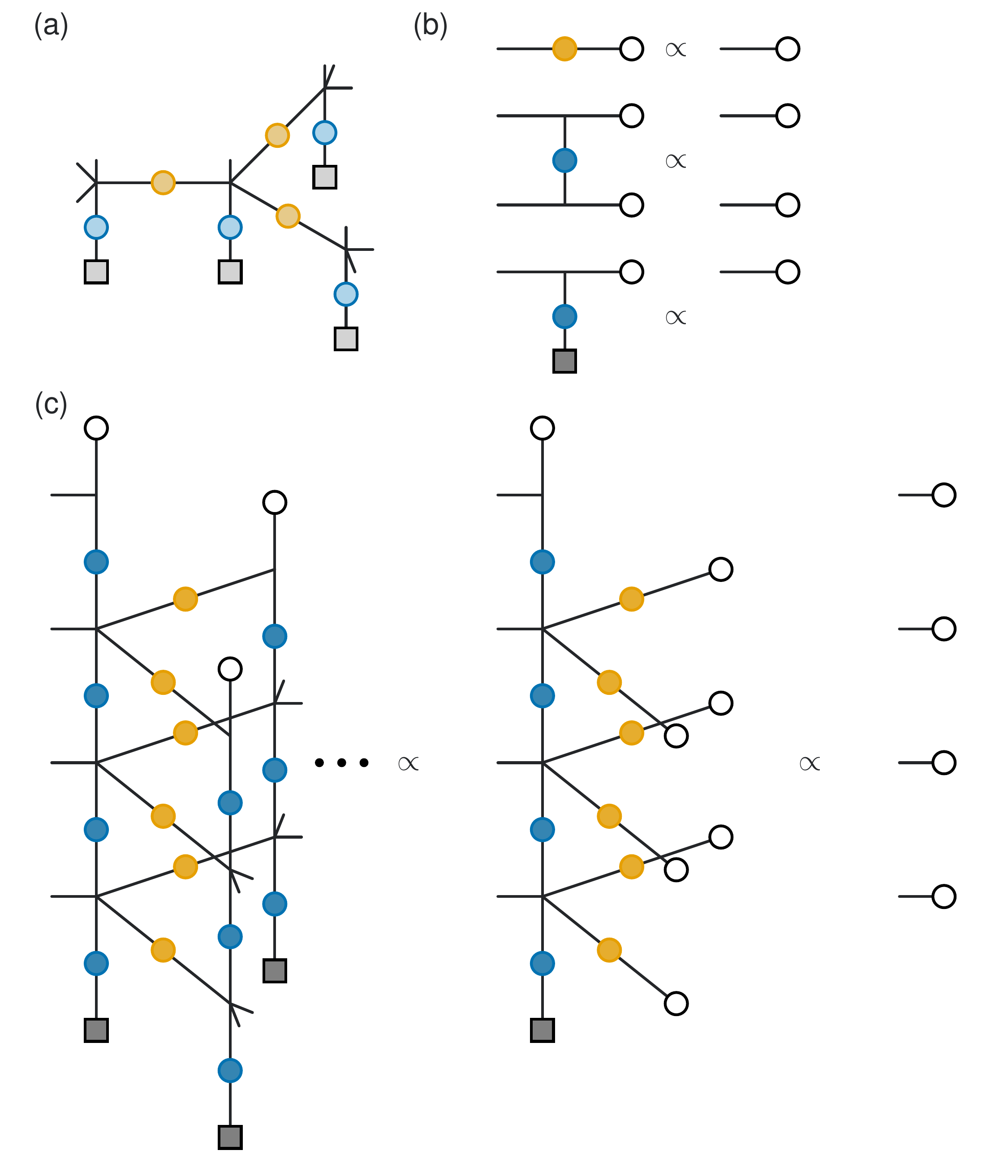}
    \caption{Tensor network diagram for the dynamics of the kicked Ising model at its perfect dephaser (PD) point. (a) Time evolution is given by a local $X$ rotation (light blue circles) and nearest neighbor $ZZ$ interaction (light orange circles). (b) A dual-unitary property of the self-dual points. White circles indicate a local trace operation. (c) Tensor network diagram of the IF from the light cone construction of the folded TN. The IF is reduced to a product form with the local trace operations recursively, where the IF satisfies the PD property. }
    \label{fig:perfect_dephaser}
\end{figure}

Belief propagation provides a numerical scheme to obtain the MPS approximation of IF. Even though an analytical expression for the IF cannot be obtained in general, Ref.~\cite{Lerose2021} discusses a class of dynamics on 1D lattices where analytical expressions can be obtained, called the perfect dephaser (PD) point, where the IF has a product form at each time that effectively dephases the system. This section provides examples of the PD point IF on tree lattices from dynamics within the family of kicked Ising models.

We start from a kicked Ising model described by the following Trotterized time evolution,
\begin{equation}
    \hat{U} = \prod_{\langle j, k \rangle} e^{i \theta_{J} \hat{Z}_j \hat{Z}_k / 2} \prod_j e^{-i \theta_h \hat{X}_j / 2}, \label{eq:kickedIsing}
\end{equation}
where $\langle j, k \rangle$ denotes the nearest neighbor pair on tree lattices. We will consider a simple product state as an initial state, $\ket{\psi_0} = \otimes_i \ket{0_i}$, and application of $\hat{U}$ for $m$ Trotter steps, $\ket{\psi(m)} = \hat{U}^m \ket{\psi_0}$. Fig.~\ref{fig:perfect_dephaser}a shows a graphical tensor network diagram of this time evolution. In the 1D kicked Ising model, the IF satisfies the PD property at $|\theta_{J}| = \pi / 2$ and $|\theta_h| = \pi / 2$, which coincides with the self-dual points of the kicked Ising model~\cite{Akila_2016, Bertini2019}. Each gate at the self-dual points satisfies a dual-unitary property~\cite{Bertini2019}, where its diagrammatic representation is illustrated in Fig.~\ref{fig:perfect_dephaser}b. White circles in the figure indicate a local trace operation. We also note that the gates at the given parameters are Clifford.

We show that these properties hold for Bethe lattices. We prove the PD property from the light cone construction of the folded TN to build the IF in Fig.~\ref{fig:perfect_dephaser}c. The consequent IF has a product form of local trace operations. The proof is based on the recursive structure of the tree. The light-cone IF with time $n$ ($n=4$ in Fig.~\ref{fig:perfect_dephaser}c) has subtrees with time $n-1$. For the recursion, assuming the local trace operation form of IF at time $n-1$, the local trace form of IF at time $n$ is derived from the dual-unitary property (Fig.~\ref{fig:perfect_dephaser}c).
A broader class of models also leads to a PD form of the IF if the constituting circuits satisfy the dual-unitary property or a generalization, such as in tri-unitary quantum circuits~\cite{Jonay2021, breach2025solvablequantumcircuitstree1}.

In the case of the kicked Ising model on loopy lattices, the local self-consistent BP equation reduces the problem to the Bethe lattice, resulting in the exact product form of the IF-MPS, which completely dephases the local density operator. 
The local reduced density operator undergoing the PD point kicked Ising dynamics also becomes an infinite-temperature thermal ensemble $\hat{\rho} \propto \proj{0} + \proj{1}$, thus agreeing with the analysis from the BP approximation.

\section{Dynamics on heavy-hex lattice}\label{sec:sec5}

\subsection{Kicked Ising model on heavy-hex lattice}

\begin{figure}[t]
    \centering
    \includegraphics[width=0.85\columnwidth]{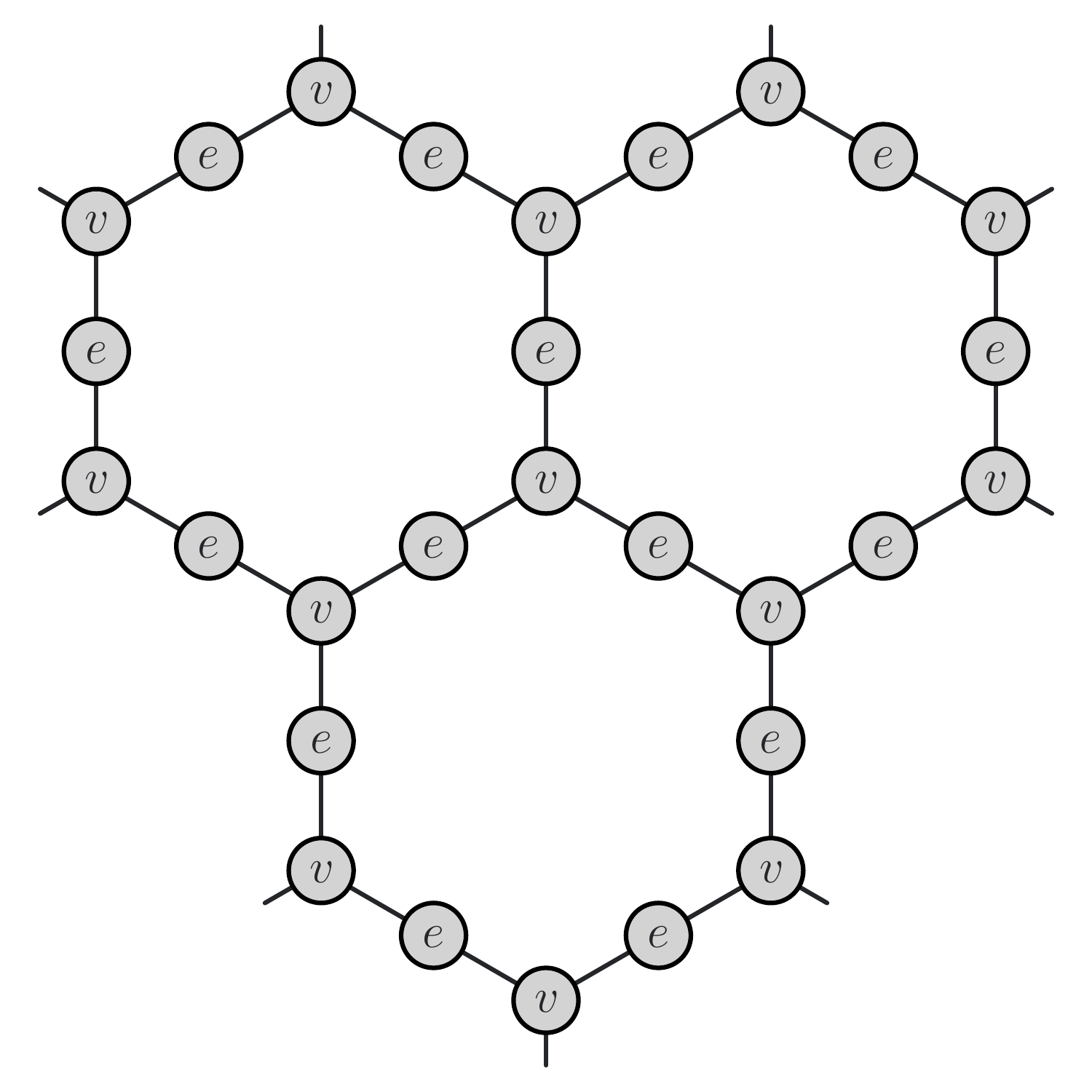}
    \caption{Geometry of a heavy-hex lattice. Qubits are placed at the vertices ($v$) and edges ($e$) of a hexagonal lattice.}
    \label{fig:heavyhex}
\end{figure}

We now study the dynamics of the kicked Ising model on an infinite heavy-hex lattice, where the qubits are placed on both the vertices and edges of a hexagonal lattice (Fig.~\ref{fig:heavyhex}). We denote each site as a $v$ site or $e$ site for the vertex and edge qubits, respectively. The time evolution operator for the kicked Ising model is given by Eq.~\ref{eq:kickedIsing}. A recent experimental study of the dynamics on this lattice using a quantum processor~\cite{Kim2023evidence} has led to the development of various numerical methods for the classical simulation of the dynamics~\cite{Begusic2024, Tindall2024ibm, liao2023simulation, anand2023classical, Kechedzhi_2024, shao2023simulating, rudolph2023classicalsurrogate, Orus2024ibm}. 

In this section, we compare our results with those obtained by time-evolving projected entangled-pair states (PEPS) or operators (PEPO) combined with belief propagation~\cite{Begusic2024, Tindall2024ibm, pancotti2023onestep, Tindall2023gauging}. In this context, belief propagation is employed to determine the gauge of the tensor network states or operators at a fixed time, or to compute expectation values after the time evolution is complete. Previous studies have shown that the BP approximation accurately described local observables and that spatial loop correlations were negligible in the heavy-hex lattice. Nonetheless, the required bond dimensions to achieve converged results were large in a few parameter regimes, and even larger bond dimensions are expected for deeper circuit simulations due to entanglement growth. These characteristics make the system a suitable testbed for the IF-BP method. 

\subsection{Numerical results}

\subsubsection{Numerical details}

Away from the perfect dephaser point, the fixed-point IF-MPS is obtained through iterative propagation using Eq.~\ref{eq:ifbp}. In this section, we assume translational and rotational invariance across the $v$ and $e$ sites in Fig.~\ref{fig:heavyhex}. The $ZZ$-type interaction terms in the folded TN (orange circles in Fig.~\ref{fig:perfect_dephaser}b) act on pairs of sites, resulting in a structure that differs from that shown in Fig.~\ref{fig:tree1}b and \ref{fig:tree1}c. We decompose the two-site operators by SVD, splitting them into single-site contributions. The square roots of the singular values,  $s^{1/2}$, are symmetrically absorbed into the adjacent sites. This decomposition yields a TN with the same geometry as in Fig.~\ref{fig:tree1}b and \ref{fig:tree1}c \footnote{Another possible representation is to absorb the $ZZ$ interaction tensors to the edge site tensors, which has been utilized in \cite{Lerose2021, SONNER2021168677}. The representation of the interaction tensor can affect the numerical results. In the continuous-time limit, the second representation suffers from zero temporal entanglement entropy~\cite{SONNER2021168677}, whereas the SVD-based representation does not~\cite{park2024tensor}. When $|\theta_J|= \pi / 2$, the singular value spectrum is uniform, and both schemes yield the same result. When $|\theta_J| \neq \pi/2 $, both representations will have the same result in the limit of the infinite bond dimension but have different numerical results in the finite bond dimension.}.

The IF-MPS is now characterized by two MPSs, $\mathcal{I}^{(t)}_{v \rightarrow e}$ and $\mathcal{I}^{(t)}_{e \rightarrow v}$. The update rule from Eq.~\ref{eq:ifbp} becomes,
\begin{equation}
    \mathcal{I}^{(t+1)}_{v \rightarrow e} = \mathcal{T}_v \cdot \left( \mathcal{I}^{(t)}_{e_1 \rightarrow v} \otimes \mathcal{I}^{(t)}_{e_2 \rightarrow v} \right),
\end{equation}
where $\mathcal{T}_v$ denotes the grouped tensor network at site $v$ and the sites $e_1$ and $e_2$ are two different incoming edge sites. Even though we distinguish $e_1$ and $e_2$, the IF-MPSs $\mathcal{I}^{(t)}_{e_1 \rightarrow v}$ and $\mathcal{I}^{(t)}_{e_2 \rightarrow v}$ are actually the same IF-MPS $\mathcal{I}^{(t)}_{e \rightarrow v}$, due to the assumption of rotational symmetry.  We can apply the same update rule for $\mathcal{I}^{(t+1)}_{e \rightarrow v}$ from $\mathcal{I}^{(t)}_{v \rightarrow e}$, but we observed that the following update rule with $\mathcal{I}^{(t+1)}_{v \rightarrow e}$ has a better convergence,
\begin{equation}
    \mathcal{I}^{(t+1)}_{e \rightarrow v} = \mathcal{T}_e \cdot \mathcal{I}^{(t+1)}_{v \rightarrow e}.
\end{equation}
This modified update rule can be understood from the iterative IF-MPS propagation in the corresponding Bethe lattice with alternating coordination numbers, $z=2$ and $z=3$.

Another important numerical detail is the initial guess for the IF-MPS, $\mathcal{I}^{(0)}_{e \rightarrow v}$. Ref.~\cite{SONNER2021168677, Lerose2023overcome} reported that the IF-MPS propagation could encounter a high (temporal) entanglement barrier depending on the initial guess, even if the entanglement of the final fixed-point IF-MPS is low. For a moderate bond dimension, we choose the initial guess IF-MPS to be the fixed point IF-MPS of the 1D lattice kicked Ising model with the same Hamiltonian parameters~\footnote{We followed the light cone tensor network scheme on the 1D system for the initial guess.}. For larger bond dimensions, we use the fixed-point IF-MPS of the lower bond dimension as an initial guess. These initial guesses converge to the fixed-point IF-MPS in a few iterations.

For numerical stability, we normalize the IF-MPS with norm 1, $\norm{\mathcal{I}^{(t)}}^2 = 1$ after each propagation step. We iterate the update until we satisfy the convergence criterion, $1 - \left|\mathcal{I}^{(t)*}_{v \rightarrow e} \cdot \mathcal{I}^{(t+1)}_{v \rightarrow e}\right| < \epsilon$ with $\epsilon = 10^{-6}$.
When we compute the expectation value of local observables $\langle \hat{O} \rangle$, we normalize the expectation value so that its total trace is 1, i.e., $\langle \hat{\mathbb{I}} \rangle = 1$, where $\hat{\mathbb{I}}$ is an identity operator. 

\begin{figure}[t]
    \centering
    \includegraphics[width=\columnwidth]{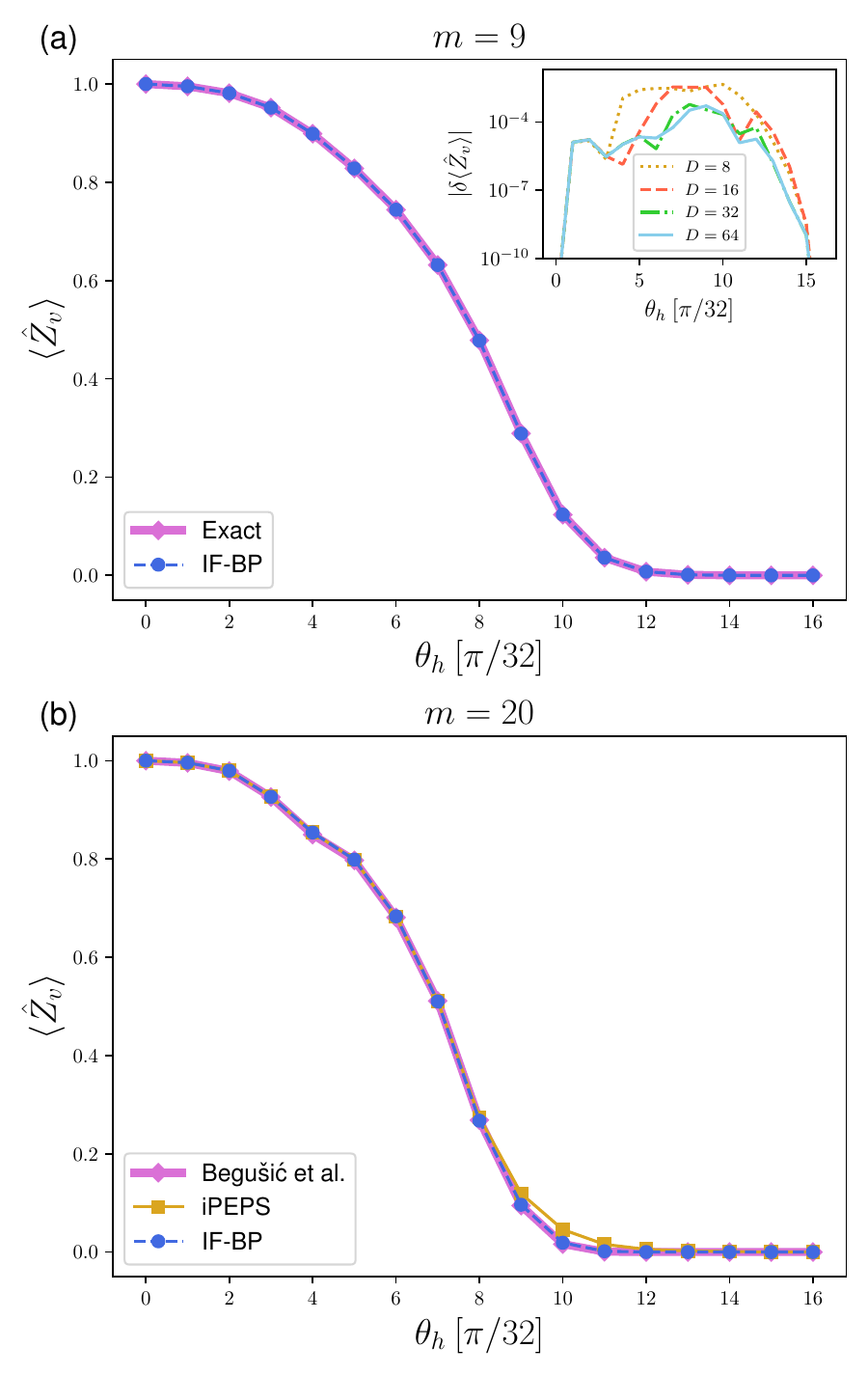}
    \caption{Magnetization on a $v$ site $\langle \hat{Z}_v \rangle$ at Trotter step (a) $m=9$ and (b) $m=20$ while varying $\theta_h$. (a) At $m=9$, the IF-BP results (blue circles) are compared to the exact benchmark from \cite{Begusic2024} (thick pink line). The inset shows the errors with respect to the exact benchmark at different bond dimensions. (b) At $m=20$, the IF-BP results are compared to the MIX TN method results from \cite{Begusic2024} (thick pink line) and iPEPS results (yellow squares).  }
    \label{fig:result1}
\end{figure}

\begin{figure*}[t]
    \centering
    \includegraphics[width=\textwidth]{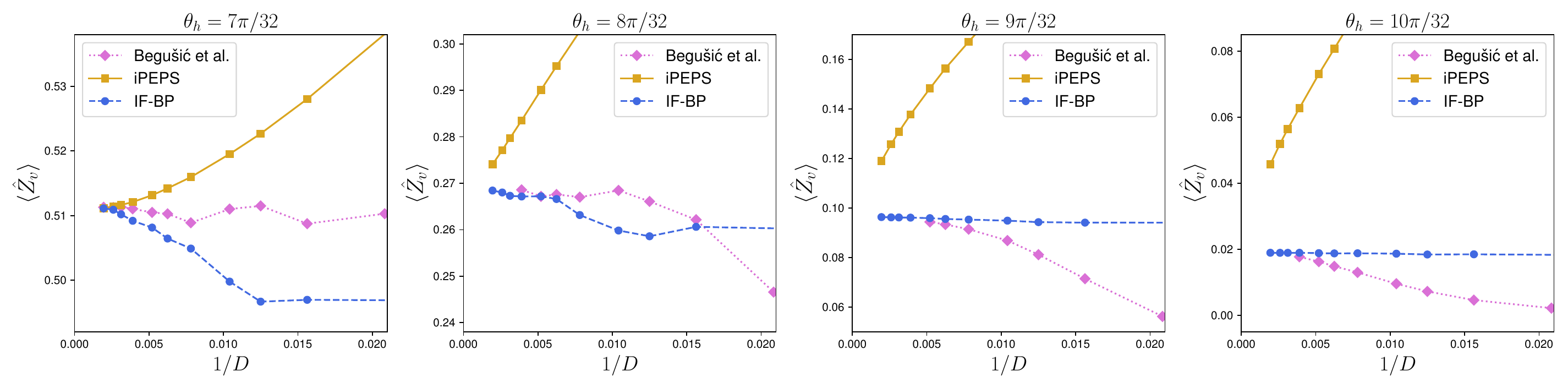}
    \caption{The convergence of magnetization $\langle \hat{Z}_v \rangle$ at $m=20$ Trotter steps with respect to bond dimension at $\theta_h = 7,8,9,10 \pi / 32$ with the MIX TN (pink dotted), iPEPS (yellow solid), and IF-BP (blue dashed) methods.}
    \label{fig:result2}
\end{figure*}

\subsubsection{Shallow depth circuit results}

We fix $\theta_{J} =  \pi/2$ and vary $\theta_h$ between zero and $\pi / 2$ with $\theta_h = k \pi / 32,$ $k \in [0,1,2, \cdots ,16]$ following the original experiment. First, we compute the magnetization on the $v$-sites, $\langle \hat{Z}_v (m) \rangle$, at Trotter steps $m=9$ and $m=20$, as shown in Fig.~\ref{fig:result1}. The $m=9$ results are compared with exact benchmarks after light cone cancellation~\cite{Kim2023evidence, Begusic2024} (Fig.~\ref{fig:result1}a). At this Trotter step $m=9$, the maximum error is less than $5 \times 10^{-3}$ using a very small bond dimension for the IF-MPS, $D=8$, and smaller than $6 \times 10^{-4}$ for bond dimension $D=32$.

We now discuss the $m=20$ results, for which no exact benchmark is available. We compare our result with those from infinite PEPS (iPEPS) calculations with the BP approximation and the results from Ref.~\cite{Begusic2024}, which employed the `MIX TN' method. In that method, half of the time evolution is carried out using a PEPS, and the other half using a PEPO. Ref.~\cite{Begusic2024} reported that the MIX TN method yields the most accurate results and the fastest convergence with bond dimension among the methods considered. While their results are based on simulations of a finite 127-qubit system, we assume the finite-size effects to be small~\cite{Tindall2024ibm, Kechedzhi_2024}.
Fig.~\ref{fig:result1}b compares the results from the IF-BP with $D=320$, iPEPS with $D=512$, and the best results from the MIX TN in \cite{Begusic2024}. Across most values of $\theta_h$, all three methods show good agreement, except for a slight deviation of the iPEPS results for $\theta_h = 9,10,11 \pi / 32$.

In Fig.~\ref{fig:result2}, we analyze the convergence behavior of the three methods with respect to bond dimension at $\theta_h = 7,8,9,10 \pi / 32$. At $\theta_h = 7\pi / 32$, all three methods converge without the need for bond dimension extrapolation. However, for larger values of $\theta_h$, the iPEPS results are not as fully converged at $D=512$ as those obtained from IF-BP and MIX TN, which accounts for the discrepancy observed in Fig.~\ref{fig:result1}. Notably, at $\theta_h = 9 \pi /32$ and $10 \pi / 32$, the IF-BP result exhibits faster convergence than both iPEPS and MIX TN methods.

\begin{figure}
    \centering
    \includegraphics[width=0.9\columnwidth]{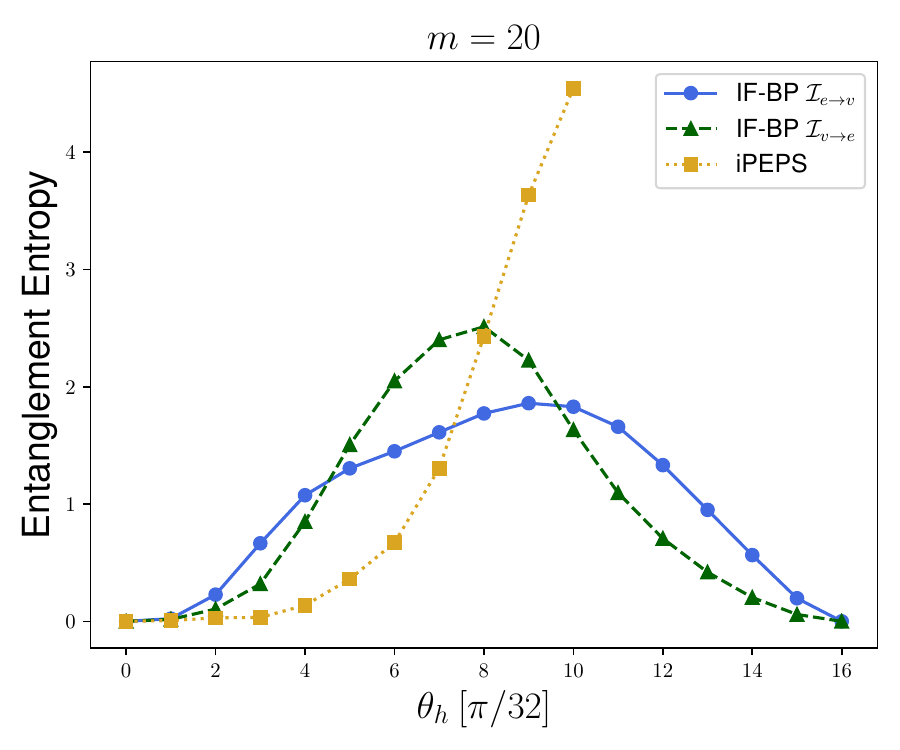}
    \caption{Entanglement entropy from the IF-BP and iPEPS methods varying $\theta_h$ at Trotter step $m=20$. The IF-BP curves show the temporal entanglement entropy of two different IF-MPS, $\mathcal{I}_{e \rightarrow v}$ (blue circles) and $\mathcal{I}_{v \rightarrow v}$ (green diamonds). The iPEPS curve (yellow square) refers to the entanglement entropy computed at each bond.}
    \label{fig:result3}
\end{figure}

We analyze the above convergence behavior by estimating the relevant entanglement entropy (EE) from the IF-BP and iPEPS methods in Fig.~\ref{fig:result3}. For the IF-BP method, we compute the EE of IF-MPS, known as the temporal entanglement entropy (TEE)~\cite{Lerose2021, SONNER2021168677}. We have two TEEs because there are two different IF-MPSs, $\mathcal{I}_{e \rightarrow v}$ and $\mathcal{I}_{v \rightarrow e}$. For the iPEPS method, we compute the EE at each bond of the iPEPS~\cite{Tindall2024ibm}. The EE of the iPEPS increases at larger $\theta_h$, whereas the TEE of IF-MPS decreases for $\theta_h>8 \pi / 32$. The IF-MPS with $\theta_h = \pi /2$ corresponds to an IF-MPS at the perfect dephaser point where the TEE is zero, which explains the above TEE trend and supports the fast convergence of the IF-BP method at $\theta_h = 9 \pi / 32$ and $\theta_h = 10 \pi / 32$. It is worth mentioning that a larger EE does not necessarily indicate lower accuracy, especially when comparing the EE of iPEPS and the TEE of IF-MPS. For example, at $\theta_h = 7 \pi /32$, the EE of iPEPS is lower than the TEE of IF-MPS, but the convergence behaviors of the two methods are similar~\footnote{Ref.~\cite{Foligno2023} analyzes the Schmidt value spectrum of the IF-MPS obtained from 1D lattice dynamics. The Schmidt value spectrum comprises a few slowly decaying large Schmidt values and a long exponential tail of Schmidt values. The TEE from this Schmidt value spectrum is often larger than the EE from a purely exponentially decaying spectrum of  Schmidt values.}.

\begin{figure}[t]
    \centering
    \includegraphics[width=0.9\columnwidth]{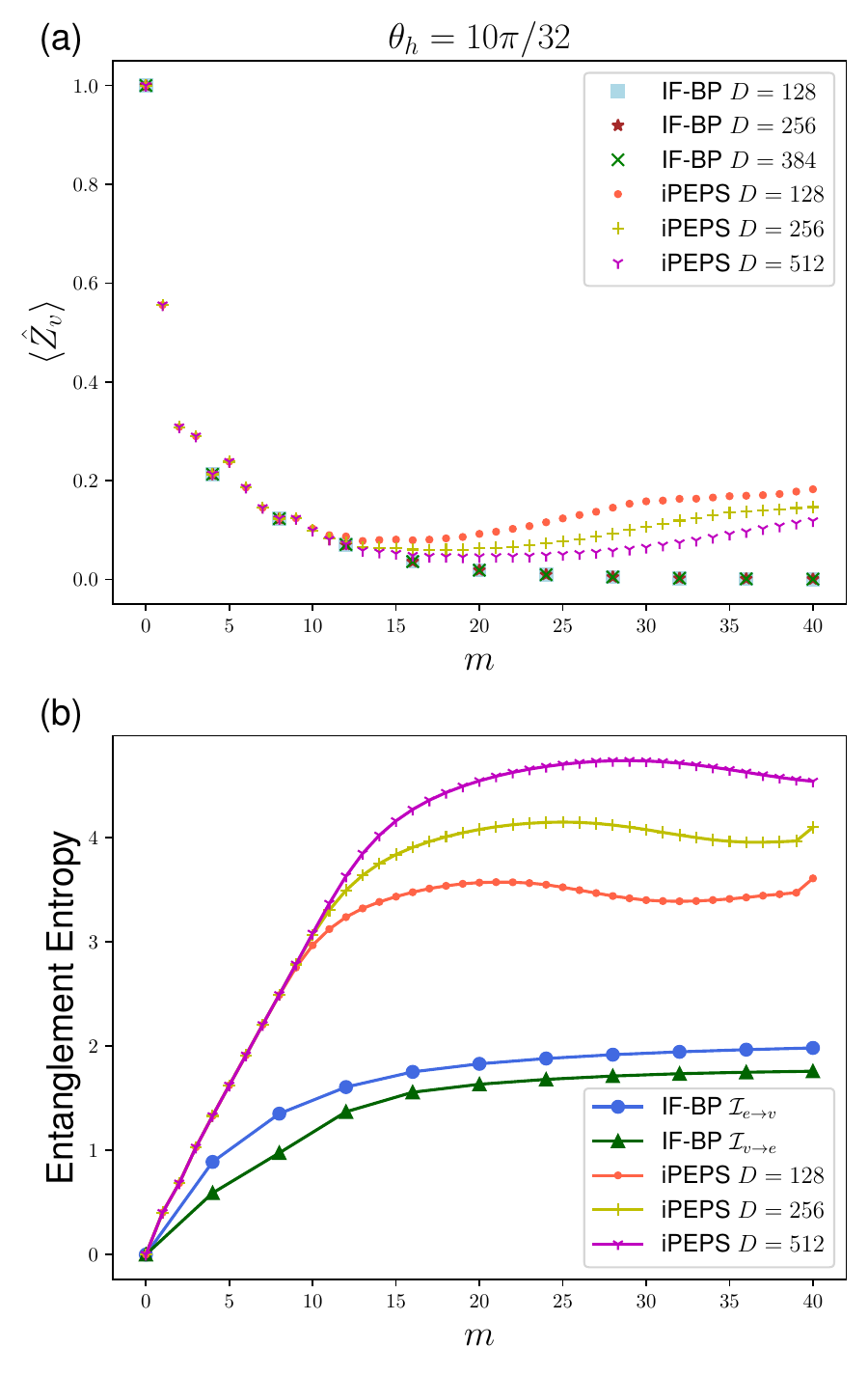}
    \caption{Time-dependence of (a) magnetization $\langle \hat{Z}_v \rangle$ and (b) entanglement entropy at $\theta_h = 10 \pi / 32$ from the IF-BP and iPEPS methods with various bond dimensions. The entanglement entropy from the IF-BP method in (b) is a converged result with respect to the bond dimension. }
    \label{fig:result4}
\end{figure}

\subsubsection{Longer time dynamics}

In this section, we present numerical results for longer-time dynamics. The EE of the wavefunction generally grows linearly with time, and TN wavefunction methods, including the iPEPS method, require exponentially growing bond dimensions. Fig.~\ref{fig:result4} shows the time-dependence of the magnetization $\langle \hat{Z}_v \rangle$ and EE at $\theta_h = 10 \pi / 32$ from the IF-BP and iPEPS methods with various bond dimensions up to Trotter step $m=40$. The magnetization from the IF-BP method already produces a converged result by a smaller bond dimension of $D=128$. The magnetization reaches a near-zero value corresponding to an infinite-temperature thermal ensemble from the Floquet thermalization~\cite{Lazarides2014, DAlessio2014, PONTE2015196}. In contrast, the iPEPS simulation shows an unphysical magnetization increase in the long-time limit. The unphysical behavior can be traced back to the plateau in the EE at long times ($m \gtrsim 20$) due to the limited bond dimension, which does not reproduce the expected linear growth of EE with time. The TEE of the IF-BP in Fig.~\ref{fig:result4}, on the other hand, is fully converged with respect to the bond dimension and does not grow linearly but grows logarithmically with time.

\begin{figure}[t]
    \centering
    \includegraphics[width=0.9\columnwidth]{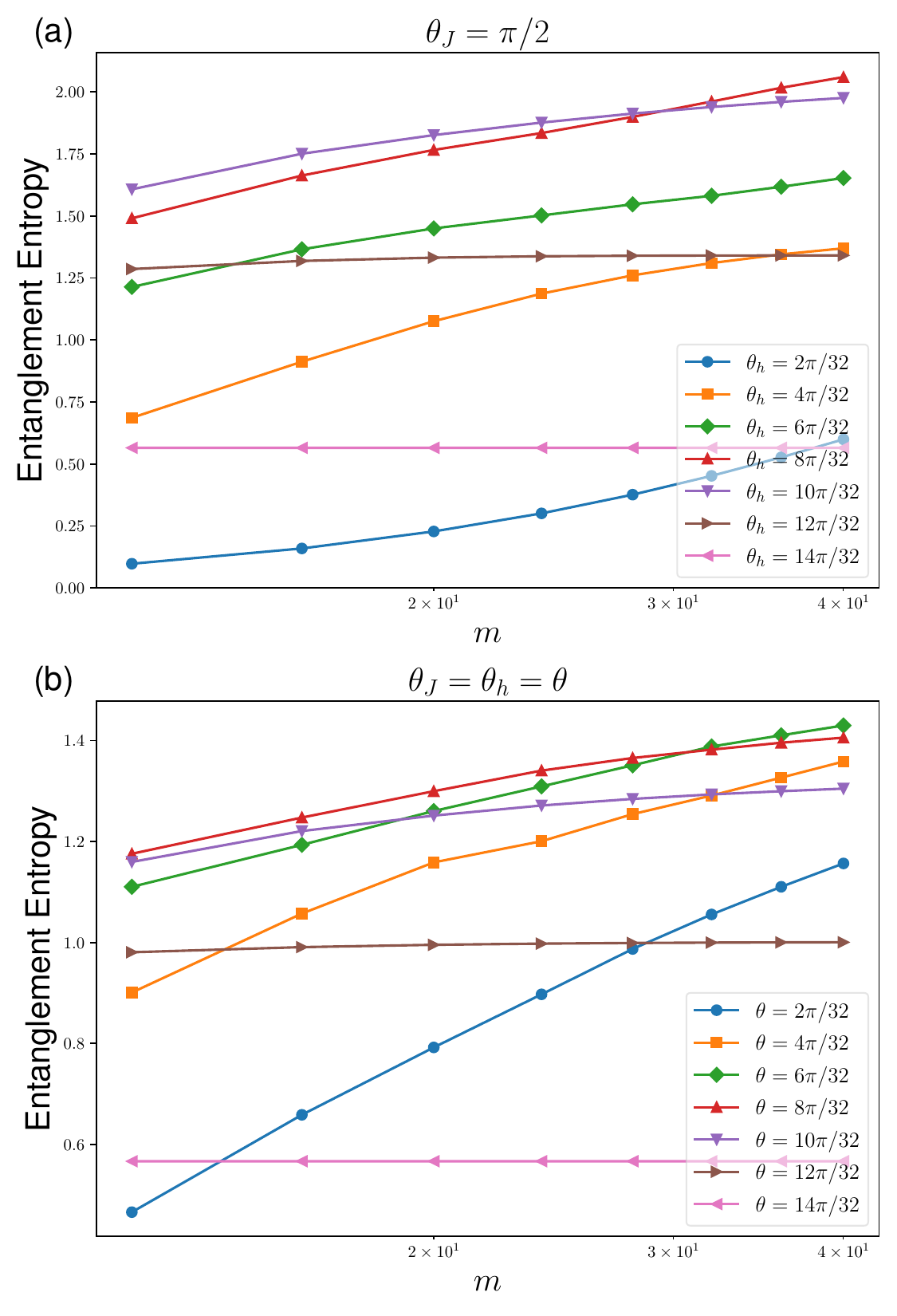}
    \caption{Logarithmic increase of temporal entanglement entropy in the IF-BP method for the IF-MPS $\mathcal{I}_{e \rightarrow v}$ when (a) varying $\theta_h$ with fixed $\theta_J$ and (b) varying $\theta_J = \theta_h = \theta$. The $m$ axis uses a logarithmic scale.}
    \label{fig:result5}
\end{figure}

In Fig.~\ref{fig:result5}, we illustrate that the logarithmic increase of the TEE holds in various parameter regimes by varying $\theta_h$ at fixed $\theta_J = \pi /2$ (Fig.~\ref{fig:result5}a) and by varying $\theta_J = \theta_h = \theta$ (Fig.~\ref{fig:result5}b). In this calculation, we use bond dimension $D=128$, based on the converged behavior of the IF-BP TEE in Fig.~\ref{fig:result4}, and show the TEE of the IF-MPS $\mathcal{I}_{e \rightarrow v}$. 

\subsubsection{Kicked Ising model with additional longitudinal field}

Previous studies in \cite{Lerose2021, Lerose2021scaling, Giudice2022, Thoenniss2023nonequilibrium} have related the logarithmic increase of TEE to the integrability. In particular, the kicked Ising model with a transverse field in a 1D lattice is exactly solvable after a Jordan-Wigner transformation to a free fermion model and shows an area-law or logarithmic behavior of TEE~\cite{Lerose2021scaling, Thoenniss2023nonequilibrium}. Unlike the 1D model, the Jordan-Wigner transformation does not map the kicked Ising model to a free-fermionic model on the heavy-hex lattice (or the Bethe lattice). Nonetheless, the observation in the previous section suggests that the TEE still grows logarithmically in this model up to the Trotter step $m=40$.

\begin{figure}[t]
    \centering
    \includegraphics[width=0.9\columnwidth]{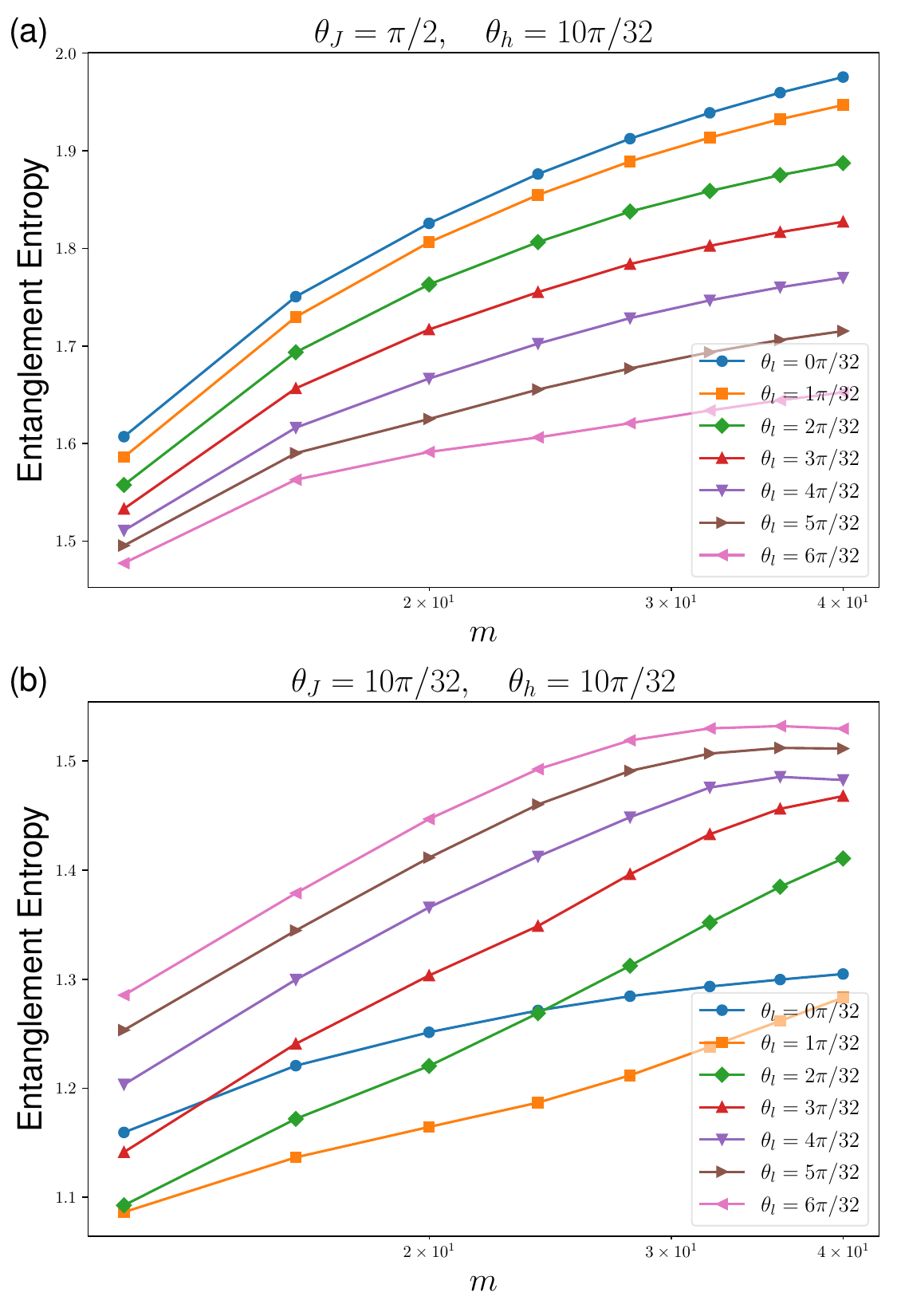}
    \caption{Temporal entanglement entropy scaling after adding the longitudinal field at (a) $\theta_J = \pi / 2$, $\theta_h = 10 \pi / 32$, and (b) $\theta_J = 10 \pi / 32$, $\theta_h = 10 \pi / 32$. The strength of the longitudinal field is controlled by $\theta_l$. The $m$ axis uses a logarithmic scale.}
    \label{fig:result6}
\end{figure}

The integrability of the 1D kicked Ising model is broken by adding a longitudinal field. We consider the following Trotterized time evolution with a longitudinal field:
\begin{equation}
    \hat{U} = \prod_{\langle j, k \rangle} e^{i \theta_{J} \hat{Z}_j \hat{Z}_k / 2} \prod_j e^{-i \theta_h \hat{X}_j / 2} e^{-i \theta_l \hat{Z}_j / 2},
\end{equation}
where $\theta_l$ controls the longitudinal field strength. Ref.~\cite{Lerose2021, Lerose2021scaling} report a regime where the TEE grows linearly with time in the kicked Ising model with a longitudinal field. Fig.~\ref{fig:result6} shows the TEE behavior varying $\theta_l$ at (a) $\theta_J = \pi / 2$, $\theta_h = 10 \pi / 32$, and (b) $\theta_J = 10 \pi / 32$, $\theta_h = 10 \pi / 32$ for kicked Ising dynamics on the heavy-hex lattice. The TEE behavior in the heavy-hex lattice shows a logarithmic or sublogarithmic increase with time, even after adding the longitudinal field. For $\theta_j = \pi / 2$, $\theta_h = 10\pi / 32$, the TEE even decreases at larger $\theta_l$. This behavior is qualitatively different from the TEE scaling in the 1D kicked Ising model.

\section{Cluster expansion of the tensor network influence functional}\label{sec:sec6}

\subsection{Method}

The BP solution provides an accurate result when the underlying graph is locally tree-like. Nevertheless, in a general loopy graph, such as a graph over a square lattice, the BP solution gives an uncontrolled approximation, because it ignores loop correlations. Numerous efforts have been made to introduce loop corrections in both classical~\cite{Kikuchi1951, Yedidia2000, Chertkov2006loop, Montanari2005, Pelizzola2005, Parisi2006, Zhou2012, Cantwell2019pnas, Kirkley2021sciencdadvances, WangZhangPanZhang2024tn} and quantum~\cite{evenbly2024loopseriesexpansionstensor} systems.

\begin{figure}[t]
    \centering
    \includegraphics[width=1.0\columnwidth]{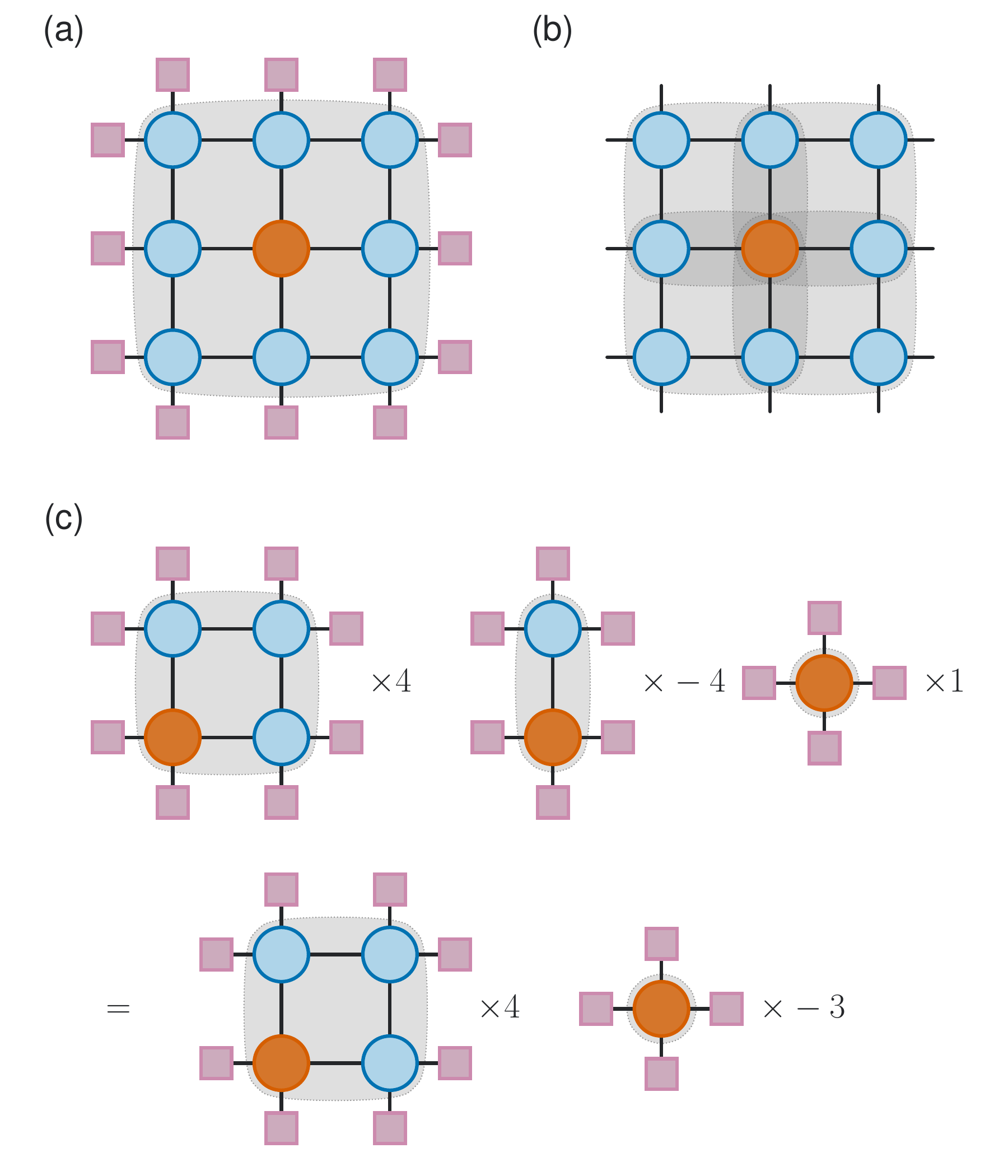}
    \caption{(a) A finite cluster around the local observable located in the center (red) with a maximum loop width $W=2$ surrounded by BP environments (pink). (b) The region with width $W=2$ can be patched together from four smaller clusters with loop width $W=1$. The cluster expansion uses the counting number of overlapping regions based on the inclusion-exclusion principle. (c) Cluster expansion up to $W=1$ includes 4-site, 2-site, and 1-site clusters with counting numbers 4, -4, and 1, respectively. The 2-site cluster with a tree geometry can be further reduced to the 1-site cluster by using the BP equations.}
    \label{fig:loop_fig1}
\end{figure}

We now describe a method to include the loop effects of quantum dynamics that are neglected in the IF-BP solution using a cluster expansion, similar to the loop series expansion presented in \cite{evenbly2024loopseriesexpansionstensor}. Direct application of \cite{evenbly2024loopseriesexpansionstensor} to the IF-BP setting is challenging due to the need to construct orthogonal message subspaces, which is complicated by the MPS structure of our messages. Instead, we choose a related approach that is more easily generalized to our setting. A detailed analysis of the cluster expansion approach is done in Ref.~\cite{gray2025tensornetworkloopcluster}, focusing on ground-state quantum many-body problems. We first consider a single cluster region around the local observable of interest and use the BP solution to define its boundary. Fig.~\ref{fig:loop_fig1}a illustrates a cluster of maximum loop width $W=2$ surrounded by its BP environment. This cluster idea has previously been used to compute local observable expectation values with PEPS wavefunction~\cite{vlaar2021, Gao2025}.

The cluster expansion combines expectation values from multiple overlapping regions to determine an improved local observable expectation value. For example, in Fig.~\ref{fig:loop_fig1}b, the region with width $W=2$ can be obtained by patching together four smaller regions with loop width $W=1$. Different regions share overlapping regions around the local observable. The counting number of the overlapping regions is evaluated based on the inclusion-exclusion principle~\cite{TangRigol2013}, including its sign to avoid double-counting. As drawn in Fig.~\ref{fig:loop_fig1}c, the cluster expansion up to $W=1$ includes 4-site, 2-site, and 1-site regions with counting numbers 4, -4, and 1, respectively. The factor of 4 reflects the rotational invariance of the dynamics.
Using the equations satisfied by the BP solution, further reductions can be made to locally tree-like regions, such as reducing the 2-site region to a 1-site region, resulting in the counting number -3 for the 1-site region. 

In general, by patching together regions with maximum loop width $W$, one can cover a region with maximum loop width $2W$, after which a counting number $c(r)$ is assigned to each region $r$ based on the inclusion-exclusion principle. This counting scheme has been exploited in many different numerical schemes, such as generalized BP~\cite{Yedidia2000} and the numerical linked-cluster expansion~\cite{TangRigol2013}. The local observable expectation can then be computed with the counting number $c(r)$, using $\langle \hat{O} \rangle_r$ to denote the expectation value from each region $r$, either by a weighted arithmetic mean,
\begin{equation}
    \langle \hat{O} \rangle \approx \sum_r c(r) \times \langle \hat{O} \rangle_r, \label{eq:expansion_add}
\end{equation}
or by a weighted geometric mean,
\begin{equation}
    \langle \hat{O} \rangle \approx \prod_r \langle \hat{O} \rangle^{c(r)}_r, \label{eq:expansion_mult}
\end{equation}
In practice, we find that the two yield numerically similar values, with differences between the two much smaller than the remaining error due to the finite loop width $W$. Henceforth, unless otherwise specified, we use the expansion with the weighted arithmetic mean in Eq.~\ref{eq:expansion_add}.

After assigning the regions, we need to compute $\langle \hat{O} \rangle_r$ by contracting the tensor network within this region. We note that the region is defined as a subgraph on the 2D lattice above. However, there is also the time direction perpendicular to the lattice, where we express the BP environment using the IF-MPS. Using exact contraction, the contraction cost scales exponentially with the number of sites in the region. Instead, we employ approximate contraction by compressing bonds between nearest neighbor sites during contractions over the time direction. We utilize the `simple update' algorithm~\cite{Jiang2008simpleupdate, Jahromi2019} for this compression, which features a bond-local gauge structure akin to that of BP compression~\cite{Tindall2023gauging, Begusic2024, Alkabetz2021}, but without implementing self-consistency over the graph. This type of simple update compression can be less accurate than BP compression due to a less self-consistent local gauge. However, we have found the loss of accuracy to be negligible here, while using this simple update reduces the computational cost by a significant factor. We propagate the tensor network with simple update compression from the initial state to the local observable and use boundary MPS contraction~\cite{verstraete2004renormalization, VerstraeteMurgCirac2008, Lubasch2014prb, Lubasch2014} for the final 2D tensor network. 
Propagation in this direction allows us to reuse tensor network intermediates for multiple regions with the same shape, but where the local observable is located in a different part of the region.

\subsection{Numerical results}

We test the performance of the cluster expansion of the TN-IF by simulating the quench dynamics of the transverse field Ising model (TFIM),
\begin{equation}
    \hat{H} = - J \sum_{\langle j, k \rangle} \hat{Z}_j \hat{Z}_k - h \sum_j  \hat{X}_j,
\end{equation}
on the 2D square lattice. Various numerical methods, such as iPEPS~\cite{Czarnik2019, Dziarmaga2021, Dziarmaga2022}, neural quantum methods~\cite{SchmittHeyl2020, Sinibaldi2024}, and sparse Pauli dynamics (SPD)~\cite{Begusic2024realtime}, have been employed to simulate the quench dynamics of the 2D TFIM. We consider quench dynamics from an initial state that is the ground state for $h \to \infty$, i.e., $\ket{\psi_0} = \bigotimes_j \ket{+}_j$ where $\ket{+} = (\ket{0}+\ket{1})/\sqrt{2}$, and use a quenched Hamiltonian with $h=h_c$ and $h=2h_c$, where $h_c = 3.04438$ corresponds to the quantum critical point~\cite{Blote2002}, and  $J=1$. For the time evolution operator $\hat{U}$ we use the second-order Trotter decomposition,
\begin{equation}
    \hat{U} = \prod_j e^{i \Delta t \hat{X}_j / 2} \prod_{\langle j, k \rangle} e^{i \Delta t \hat{Z}_j \hat{Z}_k} \prod_j e^{i \Delta t \hat{X}_j / 2}.
\end{equation}

\begin{figure}[t]
    \centering
    \includegraphics[width=1.0\columnwidth]{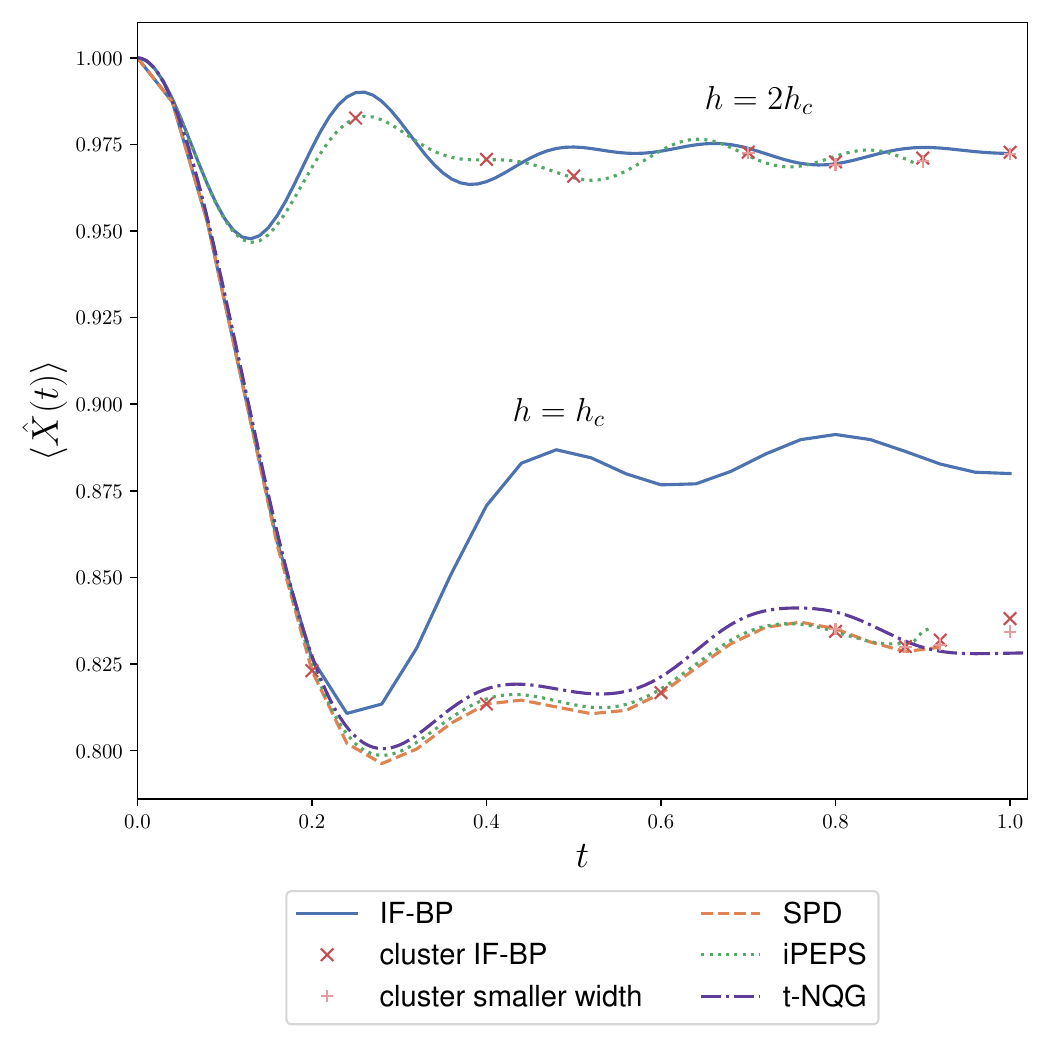}
    \caption{Time-dependence of the local observable expectation value $\langle \hat{X} (t) \rangle$ of 2D TFIM after the quench from the $X$-polarized state to $h=h_c$ and $h=2h_c$, where $h_c=3.04438$ is a critical point. We compute these dynamics using IF-BP and its cluster expansion and compare to reference data from sparse Pauli dynamics (SPD)~\cite{Begusic2024realtime}, iPEPS~\cite{Dziarmaga2022}, and time-dependent Neural Quantum Galerkin (t-NQG)~\cite{Sinibaldi2024, Sinibaldiprivate} methods. For the final four time points, results from a cluster size with one loop width smaller are included to illustrate the convergence behavior.}
    \label{fig:loop_fig_result1}
\end{figure}

Fig.~\ref{fig:loop_fig_result1} shows the time-dependence of the local observable expectation value $\langle \hat{X} (t) \rangle$ up to time $t=1.0$ from the IF-BP and its cluster expansion compared to data from the SPD~\cite{Begusic2024realtime}, iPEPS~\cite{Dziarmaga2022}, and time-dependent Neural Quantum Galerkin (t-NQG)~\cite{Sinibaldi2024, Sinibaldiprivate} methods.
The cluster IF-BP data shown in Fig.~\ref{fig:loop_fig_result1} corresponds to the largest cluster size used in our calculations; for the final four time points, results from a cluster size with one loop width smaller are also included to illustrate the convergence behavior. A more detailed analysis of the convergence with respect to cluster size is presented in a later paragraph.
We aim to compare our result to the SPD data for $h=h_c$ and the iPEPS data for $h=2h_c$. Therefore, we set the timestep $\Delta t = 0.04$ and $\Delta t = 0.01$ for the quenches with $h=h_c$ and $h=2h_c$, respectively, corresponding to the timesteps used in the SPD and iPEPS simulations. As an estimate of the corresponding time-step error, Ref.~\cite{Begusic2024realtime} reported that the Trotter timestep error at $h=h_c$ is less than $0.003$ in the observable, which agrees well with the scale of deviation between SPD and iPEPS. The t-NQG method exhibits negligible error from timestep discretization in the numerical integration.

\begin{figure}
    \centering
    \includegraphics[width=1.0\columnwidth]{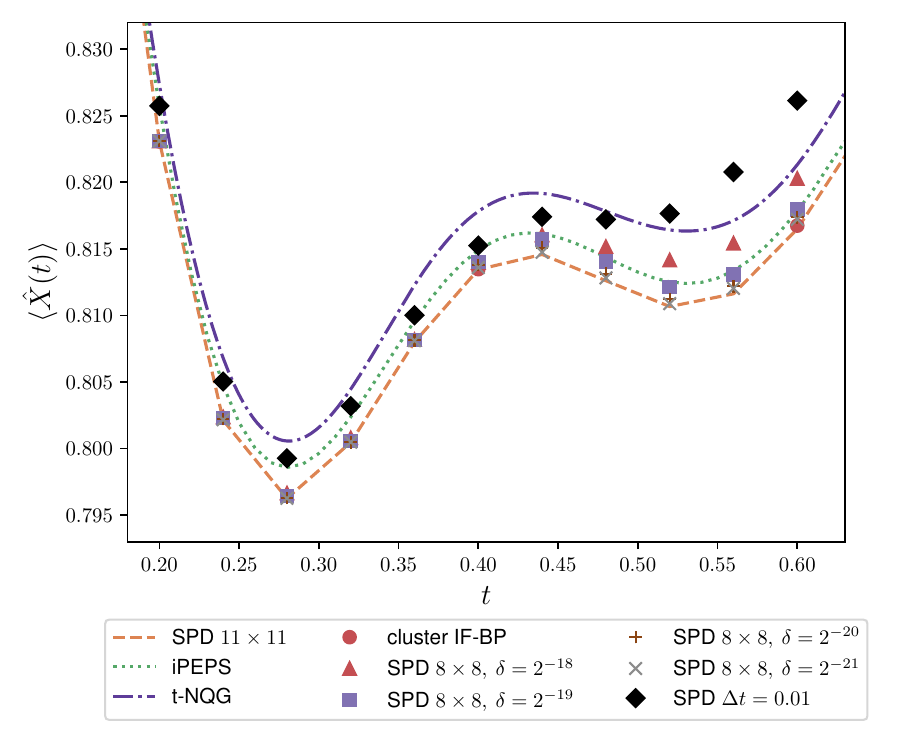}
    \caption{Finite-size analysis of the SPD method applied to an $8 \times 8$ lattice with periodic boundary conditions, matching the setup of the t-NQG method. Convergence of the SPD method is controlled by the threshold parameter $\delta$. We find that the finite size effect of the SPD method is negligible.}
    \label{fig:nqs_comparison}
\end{figure}

We note that the SPD and t-NQG methods are implemented on finite systems. Specifically, SPD is run on an $11 \times 11$ lattice with open boundary conditions, while t-NQG uses an $8 \times 8$ lattice with periodic boundary conditions. In Fig.~\ref{fig:loop_fig_result1} at $h = h_c$, some differences are observed among the reference data obtained from SPD, iPEPS, and t-NQG. To investigate the origin of these differences, we additionally performed SPD simulations on an $8 \times 8$ lattice with periodic boundary conditions, as shown in Fig.~\ref{fig:nqs_comparison} up to $t = 0.6$. The SPD method contains a controllable threshold parameter $\delta$ that enables us to verify numerical convergence. By decreasing $\delta$ from $2^{-18}$ to $2^{-21}$, we observe that the results on the $8 \times 8$ lattice converge to those from the $11 \times 11$ system, thereby ruling out finite-size effects as the source of discrepancy. Fig.~\ref{fig:nqs_comparison} also includes SPD data with $\Delta t = 0.01$ and $\delta=2^{-19}$, taken from Ref.~\cite{Begusic2024realtime}. We note that the SPD method generally requires a smaller $\delta$ to achieve comparable accuracy. Despite this, the SPD data shows good agreement with the iPEPS results up to $t = 0.4$, confirming that the visible discrepancies arise from the timestep error. Given the timestep error estimate of $0.003$ from Ref.~\cite{Begusic2024realtime} and the comparison to iPEPS data with $\Delta t = 0.01$, most of the remaining deviation of the t-NQG data likely originates from the limited variational expressivity of the particular neural quantum parametrization.

\begin{figure}
    \centering
    \includegraphics[width=1.0\columnwidth]{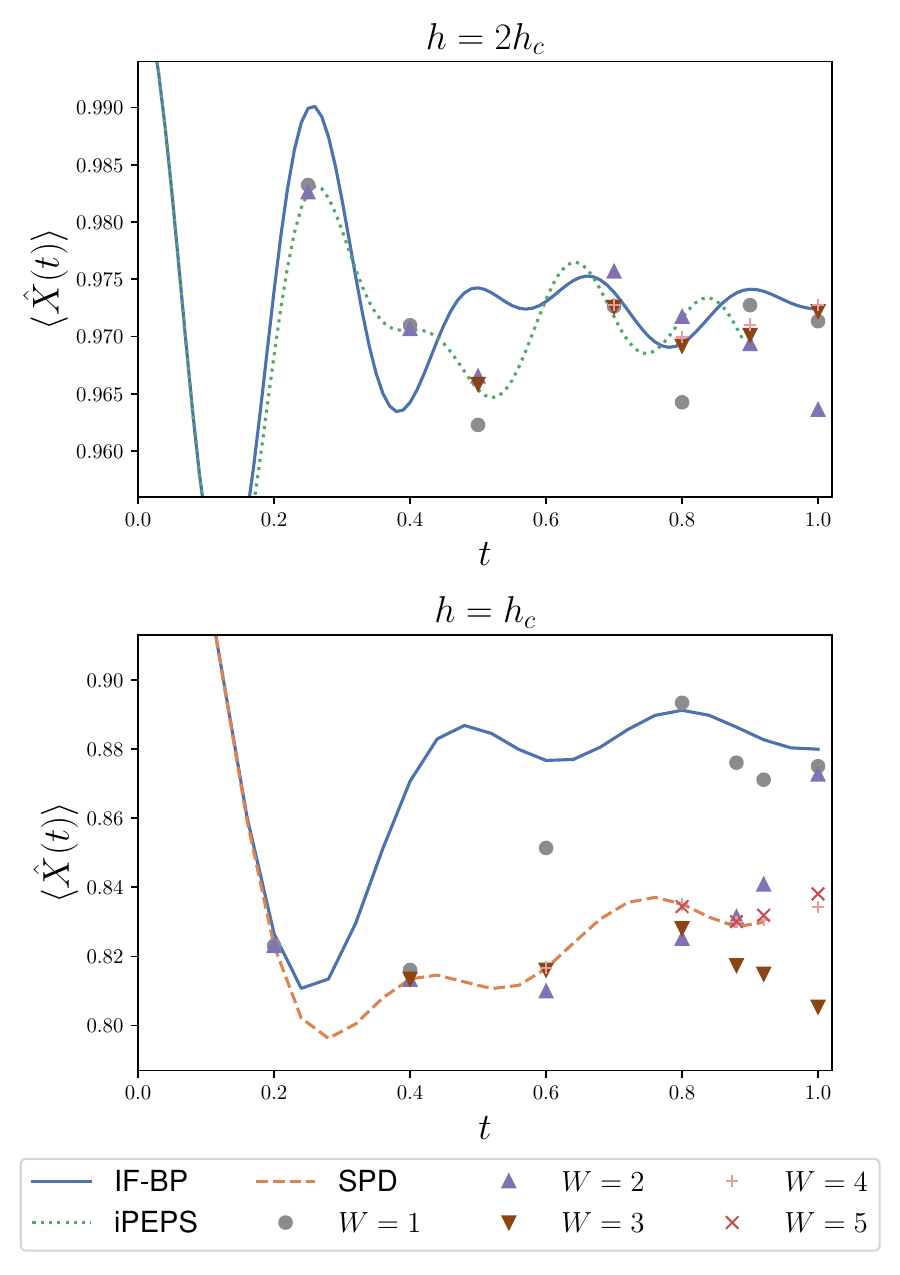}
    \caption{Convergence of $\langle \hat{X} (t) \rangle$ using a cluster expansion of IF-BP by varying maximum loop widths $W$ at $h= 2h_c$ (top) and $h=h_c$ (bottom).}
    \label{fig:loop_fig_result2}
\end{figure}

We first find the IF-BP estimate of the observable, constructing the IF-MPS through light cone propagation, as described in Eq.~\ref{eq:lightcone}. The IF-BP estimate demonstrates rapid convergence with respect to the bond dimension of the IF-MPS, achieving full convergence at $D=16$ for $t=1.0$ for both $h=h_c$ and $h=2h_c$. For $h=2h_c$, the IF-BP estimate of the expectation value provides an accurate solution, with errors of less than $0.01$ compared to that from the iPEPS simulation. In contrast, the expectation values from the IF-BP technique at $h=h_c$ accurately capture the dynamics in the short-time limit ($t < 0.2$) but exhibit noticeable deviations for $t > 0.2$. 

Next, we incorporate loop effects through the cluster expansion. In Fig.~\ref{fig:loop_fig_result2}, we show cluster IF-BP results with various maximum loop widths $W$.  For $h=2h_c$, the cluster expansion converges rapidly with respect to $W$, achieving convergence to within $0.002$ at $W=4$ and showing good agreement with the iPEPS data. However, for $h=h_c$, an accurate description requires larger $W$ for the longer-time dynamics. For instance, the results with $W=1$ and $W=2$ remain accurate up to $t=0.4$ but deviate beyond $t=0.6$. Similarly, the results with $W=3$ begin to deviate at $t=0.8$. Nevertheless, results with $W=4$ and $W=5$ show good convergence to the reference SPD data up to $t=0.92$, with errors of less than 0.002.

Another factor determining the computational cost of the cluster expansion is the intermediate bond dimension $D_{\su}$ used in the SU compression and the boundary MPS contraction of the clusters. In our calculations, the boundary MPS bond dimension was set equal to $D_{\su}$,  which provided sufficient accuracy. The required $D_{\su}$ increases with $t$ to maintain accuracy in the tensor network contraction. Specifically, the required $D_{\su}$ to converge to the SPD reference data within 0.002 at $h=h_c$ is 24, 40, 56, and 64 for $t=0.4, 0.6, 0.8$, and $0.8 < t < 1.0$, respectively. While $D_{\su}=72$ was used for $t=1.0$, this did not yield fully converged results for large $W$, resulting in an estimated uncertainty of 0.005 in the observable. Consequently, the visible difference between $W=4$ and $W=5$ at $t=1.0$ in Fig.~\ref{fig:loop_fig_result2} arises not only from loop effects but also from insufficient convergence in $D_{\su}$. The required increase in $D_{\su}$ stems from the entanglement growth within the finite cluster over time, which remains a challenge for the cluster-based classical simulations. We note that, even though the resulting $D_{\su}$ may appear large from the viewpoint of conventional time-dependent PEPS simulation, in this case it corresponds to the bond dimension in the density operator space, and thus should be compared to the square of the corresponding PEPS bond dimension. 
In comparison, the maximum bond dimension for iPEPS used in \cite{Dziarmaga2022} was $D=6$. This employed a corner transfer matrix renormalization group method~\cite{Corboz2016, CorbozRice2014} both to compute the observable expectation values and to optimize the PEPS at each time step, which limits the maximum bond dimension that can be used.

\section{Conclusions}\label{sec:sec7}

In this work, we introduced the belief propagation (BP) algorithm based on tensor network influence functionals to simulate the dynamics of local observables in quantum lattice systems beyond one dimension. The method is numerically exact on tree lattices in the limit of large bond dimensions and remains accurate on locally tree-like lattices with large loops. Our numerical results validate the effectiveness of IF-BP by reproducing the kicked Ising dynamics quantum experiments on the heavy-hex lattice. Moreover, the IF-BP approach outperforms traditional tensor network state-based methods for longer-time dynamics, as it leverages the slow, logarithmic growth of temporal entanglement entropy to maintain computational efficiency. The framework also extends naturally to computing multi-time correlation functions of local observables.

To address the limitations of BP on graphs with loops, we developed a cluster expansion of the tensor network influence functionals. By enlarging clusters around the subsystem and combining expectation values from overlapping regions, the method systematically incorporates loop-induced correlations and provides controlled improvements in accuracy beyond the BP approximation. We demonstrated the effectiveness of this approach by simulating the quench dynamics of the transverse field Ising model on a 2D square lattice, where loop effects are prominent. The cluster expansion produced results that are competitive with, and in some cases improve upon, existing state-of-the-art methods, underscoring its potential as a practical tool for simulating nonequilibrium dynamics in 2D quantum systems. In addition to advancing classical simulation techniques, such numerical methods also provide a valuable tool for benchmarking quantum devices~\cite{Kim2023evidence, King2025dwave, haghshenas2025quantinuum}.

\section{Acknowledgements}

We thank Jacek Dziarmaga, Tomislav Begu\v{s}i\'{c}, and Alessandro Sinibaldi for sharing the iPEPS~\cite{Dziarmaga2022}, SPD~\cite{Begusic2024realtime}, and t-NQG~\cite{Sinibaldi2024, Sinibaldiprivate} data, respectively, presented in Fig.~\ref{fig:loop_fig_result1}. The authors thank Tomislav Begu\v{s}i\'{c}, Giuseppe Carleo, Filippo Vicentini, and Alessandro Sinibaldi for helpful discussions.  This material is based upon work supported by the U.S. Department of Energy, Office of Science, Office of Advanced Scientific Computing Research and Office of Basic Energy Sciences, Scientific Discovery through Advanced Computing (SciDAC) program under Award Number DE-SC0022088. The \texttt{quimb} library~\cite{gray2018quimb} has been used in the numerical experiments. Computations presented here were conducted in the Resnick High Performance Computing Center, a facility supported by Resnick Sustainability Institute at the California Institute of Technology.
G.P. acknowledges support from the Eddleman Quantum Graduate Fellowship at Caltech.

\bibliography{references} 

\end{document}